\documentclass[preprint]{raa}            

\usepackage{graphicx,times}             
\usepackage{natbib}
\usepackage{amssymb,amsmath}
\usepackage{physics}
\bibpunct{(}{)}{;}{a}{}{,}

\usepackage{soul}
\usepackage{xcolor}

\newcommand{\hi}{H\textsc{i} }

\usepackage[pagebackref=true]{hyperref}
\hypersetup{
    colorlinks=true,
    linkcolor=blue,
    filecolor=magenta,
    urlcolor=blue,
    citecolor=blue,
    pdfpagemode=FullScreen,
    }

\begin{document}

  \title{Application of Regularization Methods in the Sky Map Reconstruction of the Tianlai Cylinder Pathfinder Array }
  \titlerunning{Tianlai Cylinder Imaging}

   \volnopage{Vol.0 (20xx) No.0, 000--000}      
   \setcounter{page}{1}          

    \author{Kaifeng Yu \inst{1,2,3}
    \and Shifan Zuo \inst{1,2}
    \and Fengquan Wu \inst{1,2}
    \and Yougang Wang \inst{1,2}
    \and Xuelei Chen$^*$
        \inst{1,2,3,4,5}
    }
\institute{National Astronomical Observatories, Chinese Academy of Sciences, Beijing 100101,
 China; \\
\and
Key Laboratory of Radio Astronomy and Technology, Chinese Academy of Sciences, Beijing 100101, China;\\
\and
School of Astronomy and Space Science, University of Chinese Academy of Sciences, Beijing 100049, China\\
\and
Key Laboratory of Cosmology and Astrophysics (Liaoning) \& College of Sciences, Northeastern University, Shenyang 110819, China\\
\and
Center of High Energy Physics, Peking University, Beijing 100871, China. \\
$^*${\it xuelei@cosmology.bao.ac.cn}
}

\vs\no
   {\small Received 20xx month day; accepted 20xx month day}

\abstract{The Tianlai cylinder pathfinder is a radio interferometer array to test 21 cm intensity mapping techniques in the post-reionization era. It works in passive drift scan mode to survey the sky visible in the northern hemisphere. To deal with the large instantaneous field of view and the spherical sky, we decompose the drift scan data into $m$-modes, which are linearly related to the sky intensity. The sky map is reconstructed by solving the linear interferometer equations. Due to incomplete $uv$ coverage of the interferometer baselines, this inverse problem is usually ill-posed, and regularization method is needed for its solution. In this paper, we use simulation to investigate two frequently used regularization methods,  the Truncated Singular Value Decomposition (TSVD), and the Tikhonov regularization techniques. Choosing the regularization parameter is very important for its application. We employ the generalized cross validation (GCV) method and the L-curve method to determine the optimal value. We compare the resulting maps obtained with the different regularization methods, and for the different parameters derived using the different criteria. While both methods can yield good maps for a range of regularization parameters, in the Tikhonov method the suppression of noisy modes are more gradually applied, produce more smooth maps  which avoids some visual artefacts in the maps generated with the TSVD method. 
}

   \authorrunning{K.-F. Yu, S.-F. Zuo \& X.-L. Chen }            
   \titlerunning{Sky map reconstruction of Tianlai Cylinder Pathfinder}  

   \maketitle

%
%
\section{Introduction}
\label{sec:intro}

The Tianlai experiment is designed to develop and test the \hi Intensity Mapping (IM) technique \citep{2012IJMPS..12..256C, 2020SCPMA..6329862L}, in which the large scale distribution of neutral hydrogen is observed at low angular resolutions without resolving individual galaxies. This allows fast survey speed to cover the large volume required for cosmological studies \citep{2008PhRvL.100i1303C}.
This technique has been applied to the observation of the Green Bank Telescope (GBT) and the Parkes telescope \citep{2010Natur.466..463C, 2013ApJ...763L..20M,2013MNRAS.434L..46S,2017MNRAS.464.4938W,2018MNRAS.476.3382A,2021arXiv210204946W}. Other existing or ongoing \hi IM experiments focusing on the late-time cosmology include both single-dish telescopes and interferometers, such as FAST \citep{2020MNRAS.493.5854H}, BINGO \citep{2013MNRAS.434.1239B}, CHIME \citep{2022arXiv220201242C}, and SKA in the near future \citep{2020PASA...37....7S}.
However, \hi IM observation is very challenging due to the strong foreground radiation that is $4 \sim 5$ orders of magnitude brighter than the cosmological signal. Additionally, the instrumental systematics and other observational effects also complicate the separation of the two, a highly sophisticated analysis is required to extract this signal \citep{2020PASP..132f2001L}.

The Tianlai experiment includes a cylinder array and a dish array. The Tianlai cylinder pathfinder array is fixed on the ground and relies on the rotation of the Earth to survey the sky. It consists of three adjacent $15{\rm m} \times 40{\rm m}$ cylindrical reflectors, on which there are $31$, $32$, $33$ feeds equipped from east to west, respectively. The basic performance of the Tianlai arrays has been analyzed in \citet{2020SCPMA..6329862L} for the cylinder pathfinder, and in \citet{2021MNRAS.506.3455W} for the dish pathfinder. Although making synthesis images of the sky from the interferometric raw data is strictly speaking not needed for the power spectrum estimation, in practice it is still an essential procedure to compress the data for further scientific analysis and to provide an intuitive means of checking the data quality and algorithms applied.

The output of an interferometer, the interferometric visibility, is the cross-correlation between the voltage of two feed elements. For the unpolarized case, the visibility is given by
\begin{eqnarray}
    V_{ij}(t) &=& \int T(\hat{\vb*{n}}) A_i(\hat{\vb*{n}};t) A_j^{*}(\hat{\vb*{n}};t) e^{2\pi i \hat{\vb*{n}} \cdot \vb*{u}_{ij}(t)} \dd^2 \hat{\vb*{n}},\\
    &=&  \int T(\hat{\vb*{n}}) {B}_{ij}(\vb*{n}; t) \dd^2 \hat{\vb*{n}}.
\label{eq:vis-1}
\end{eqnarray}
where $T(\hat{\vb*{n}})$ is the sky brightness temperature, $\hat{\vb*{n}}$ gives the direction on the sky, $A_i$ is the beam of feed $i$, $\vb*{u}_{ij} = (\vb*{r}_i - \vb*{r}_j) / \lambda$ is the separation between feeds in the unit of observed wavelength. In the second line, we introduce the beam transfer function $ {B}_{ij}$. In discrete form, the integral is replaced by a sum over pixelated sky indexed by $n_i$ as following
\begin{equation}
    V_{ij}(t) = \sum_{n_i} T(n_i) B_{(ij;t)}(n_i) \Delta \Omega,
    \label{eq:vis-1-2}
\end{equation}
where $\Delta \Omega$ is the pixel angular size, or in a matrix-vector form,
\begin{equation}
    \vb*{V} = \vb*{B T}.
    \label{eq:vis-1-3}
\end{equation}
Synthesis imaging is to estimate $T(\hat{\vb*{n}})$ given the measurements $V_{ij}$. In the flat sky limit, the measured visibilities of the interferometer array correspond to Fourier modes of sky intensity variation, with the orientation and angular scale determined by the direction and length of the baselines. However, in practice the baselines of an array often do not form a complete coverage of the spatial frequency domain, so reconstructing the sky map from the visibilities is not trivial, as it is mathematically an ill-posed inverse problem, with no unique solution.  To overcome the gap of measured visibilities on the $uv$ plane,  a number of techniques have been developed to reconstruct the image in such cases, e.g., the variants of CLEAN algorithm \citep{1974A&AS...15..417H}, and some burgeoning methods based on neural networks \citep{2020RAA....20..170X, 2022A&A...664A.134S, 2022MNRAS.514.2614C}.

The general principle of sky image reconstruction for Tianlai Array was investigated in \citet{2016MNRAS.461.1950Z,2016RAA....16..158Z}. \citet{ZUO2021100439} developed a pipeline for the map-making process for the Tianlai array. In a recent paper (\citealt{2023RAA....23j5008Y}, heretofore referred to as paper I), we presented a simulation of the Tianlai cylinder pathfinder array up to the making of synthesis map, taken into account both thermal noise and calibration error. This simulation reveals that although the Tianlai array is a relatively compact and dense array, there are still incompleteness in its baseline coverage, and as a result, the reconstructed image has some artifacts arising from the incomplete \textit{uv} coverage or beam side lobes. In the present paper, we shall investigate regularization methods for dealing with the ill-posed inversion problem. 

The structure of this paper is as follows. In Section \ref{sec:mmode}, we present our simulation setup and give a brief summary of the $m$-mode formalism for map-making. In Section \ref{sec:reg-solver}, we apply the regularized $m$-mode formalism imaging to the simulation data and explore the choice of appropriate regularization parameters. In Section \ref{sec:discussion}, we discuss the error in these regularized maps, and finally in Section \ref{sec:conclusion}, we present our conclusion.

\section{Map-making with $m$-modes}
\label{sec:mmode}

\subsection{The $m$-mode Formalism}
Although imaging the sky by solving for $\vb*{T}$ in Equation \eqref{eq:vis-1-3} directly is intuitive, the amount of computation is very large. If the interferometer has $N_{\rm bl}$ baselines and has run for $N_{\rm time}$ rounds, for each single frequency and $N_{\rm pix}$ discrete sky pixels, the visibility vector $\vb*{V}$ has a size of $N_{\rm bl} \times N_{\rm time}$, the dimensions of beam transfer matrix $\vb*{B}$ will be $(N_{\rm bl} \times N_{\rm time},\; N_{\rm pix})$. \cite{2017MNRAS.464.3486Z} demonstrated this method for a small array. For Tianlai cylinder pathfinder, the number of non-redundant baselines is $N_{\rm bl} \approx 3300$, and the time samples are $N_{\rm time} \approx 21600$ in a sidereal day for $4$ s integration time. If we pixelate the whole sky within latitude $[-30^{\circ}, \; 90^{\circ}]$ in HEALPix scheme with $N_{\rm side} = 256$ \citep{2005ApJ...622..759G}, the pixel number is $N_{\rm pix} \approx 6\times10^{5}$. Then the dimensions of the transfer matrix $\vb*{B}$ would be $71{\rm M} \times 0.6{\rm M}$ per frequency. Apparently, solving Equation \eqref{eq:vis-1-3} directly would be intractable for a large array, due to the large matrix involved.

The $m$-mode method \citep{2014ApJ...781...57S, 2015PhRvD..91h3514S, 2016RAA....16..158Z} provides a convenient and computationally efficient approach for the data processing and analysis of drift scan telescopes, especially wide-field telescopes. It has been applied in the data analysis of LWA \citep{2018AJ....156...32E}, EDA2 \citep{2022PASA...39...17K} and CHIME \citep{2022arXiv220201242C}. We also have implemented this method in the map-making procedure of Tianlai data processing pipeline \texttt{tlpipe} \citep{ZUO2021100439}.

As the drift scan telescope measures the sky periodically with the rotation of the Earth, we can decompose $T(\hat{\vb*{n}})$ and $B_{ij}(\hat{\vb*{n}}; \phi)$ in spherical harmonics,
\begin{align*}
    B_{ij}(\hat{\vb*{n}}; \phi) &= \sum_{lm} B^{ij}_{lm}(\phi) Y_{lm}^{*}(\hat{\vb*{n}}) \\
    T(\hat{\vb*{n}}) &= \sum_{lm} a_{lm} Y_{lm}(\hat{\vb*{n}}),
\end{align*}
the Fourier transform of the visibility $V_{ij}(\phi)$ can be written as
\begin{align}
    \nonumber
    V_{m}^{ij} &= \int \frac{d \phi}{2\pi}V_{ij}(\phi) e^{-im \phi} \\
    &= \sum_{lm'} \int \frac{d \phi}{2\pi} B^{ij}_{lm'}(\phi) a_{lm'} e^{-im \phi},
    \label{eq:vis-3}
\end{align}
and take the rotated beam transfer function $B^{ij}_{lm}(\phi) = B^{ij}_{lm}(\phi = 0) e^{im\phi}$, Equation \eqref{eq:vis-3} adding up the noise term becomes
\begin{equation}
    V_{m}^{ij} = \sum_{l} B_{lm}^{ij} a_{lm} + n^{ij}_m,
    \label{eq:m-mode1}
\end{equation}
which gives the key equation in $m$-mode analysis, we can rewrite it in matrix form as
\begin{equation}
    (\vb*{v}_{m})_{(ij, \pm)} = (\vb*{B}_{m})_{(ij, \pm)(l)} (\vb*{a}_{m})_{l} + (\vb*{n}_m)_{(ij, \pm)},
    \label{eq:mmode-2}
\end{equation}
The label $(ij, \pm)$ indicates that the positive and negative values of $m$ are grouped together for the fact that the positive and negative $m$-modes are measurements of the same real-valued sky, which gives $a_{l, -m} = (-1)^{m}a_{lm}$.
We can rewrite Equation \eqref{eq:m-mode1} in a general form as
\begin{equation}
    \vb*{v} = \vb*{Ba} + \vb*{n},
    \label{eq:mmode-3}
\end{equation}
where matrix $\vb*{B}$ has a block diagonal structure, so we can treat each $m$-block independently. Then the Fourier transform of the visibility $\vb*{v}$ is related to the spherical harmonics coefficients of the sky $\vb*{a}$ by a simple linear equation. The imaging process is to solve for $\vb*{a}$ given the observation $\vb*{v}$ with prior information about $\vb*{B}$ and available $\vb*{n}$. Compared with directly solving for $T(n_i)$ in Equation (\ref{eq:vis-1-2}), the $m$-mode method handles matrix operations with much smaller matrices, which increased the computational efficiency tremendously. The main portion of computation is devoted to the transfer matrix $\vb*{B}$. However, once computed and stored, it can be reused in subsequent processing.

\subsection{Simulation}

We then simulate the map-making process, here the unpolarized case with a single frequency (750 MHz) is considered.
We use the publicly available software \texttt{cora}\footnote{\url{https://github.com/radiocosmology/cora}}\citep{2014ApJ...781...57S,2015PhRvD..91h3514S} to generate our mock sky model, which contains foreground including diffuse emission from the Galaxy and extra-galactic point sources, alongside the cosmological 21 cm signal. The diffuse emission is generated by extrapolating Haslam map with a specified spectral index map, and including random spectral and angular fluctuations. The extra-galactic point sources consist of a catalogue of real bright sources from NVSS and VLSS, a synthetic catalogue of fainter sources, and a random background for even fainter sources. The cosmological 21 cm signal is generated by drawing Gaussian realizations of a power spectrum \citep{2014ApJ...781...57S,2015PhRvD..91h3514S}.

We model the beam pattern, which is characterized as a long strip along the North-South direction, in the following analytical form (see paper I for details), with parameters determined from a fit to the electromagnetic simulation of Tianlai cylinder array \citep{2022RAA....22f5020S}:
\begin{equation*}
    A(\hat{\vb*{n}}) = A_{\text{NS}} (\sin^{-1}(\hat{\vb*{n}} \cdot \hat{\vb*{x}}); \theta_{\text{NS}}) \times A_{\text{EW}} (\sin^{-1}(\hat{\vb*{n}} \cdot \hat{\vb*{y}}), F, \theta_{\text{EW}}),
\end{equation*}
where $\hat{\vb*{x}}$ and $\hat{\vb*{y}}$ are the unit vector pointing East and North, respectively, and
\begin{equation}
    A_{\text{NS}}(\theta; \nu) = \exp[-4 \ln{2} \left( \frac{\theta}{\theta_{\text{NS}}(\nu)} \right)^2],
    \label{eq:beam-ns}
\end{equation}
\begin{equation}
     A_{\text{EW}}(\theta;F, \theta_{\text{EW}})
     \propto \int^{\frac{1}{4F}}_{-\frac{1}{4F}} A_D\left(2 \tan^{-1} u ;\theta_\text{EW}\right) e^{- i\frac{4\pi F W u}{\lambda} \sin \theta} \dd u,
    \label{eq:beam-ew}
\end{equation}
where $\theta_{\text{NS}}(\nu) = \alpha \frac{\lambda}{D_{\text{NS}}}$, in which $D_{\text{NS}} = 0.3$ m as the size of the Tianlai cylinder feeds, and $A_D$ is taken as
\begin{equation*}
    A_{D}(\theta; \theta_{\text{FWHM}}) = \exp \left[ -\frac{\ln 2}{2} \frac{\tan^2\theta}{\tan^2(\theta_{\text{FWHM}}/2)} \right].
\end{equation*}
We use the mean of the power beam of X and Y polarization at $750$ MHz from \citet{2022RAA....22f5020S} to fit the models above. The fitting parameters are $\alpha = 1.04$, $F = 0.2$, $\theta_{\text{EW}} = 2.74$.

The noise in each $m$-mode \citep{2015PhRvD..91h3514S} is generated by sampling a Gaussian distribution with RMS
\begin{equation}
    \delta_{ij} = \frac{\Omega_{ij} \sqrt{T_{i}^{\text{sys}} T_{j}^{\text{sys}}}}{ \sqrt{N_\text{day} N_\text{red} t_\text{sid} \Delta \nu}} \mathrm{sinc} \left(\pi\frac{m t_{\text{int}}}{t_\text{sid}}\right),
    \label{eq:noise_rms}
\end{equation}
where $T_{i}^{\text{sys}}$ and $T_{j}^{\text{sys}}$ are the system temperature for feed $i$ and $j$, $\Omega_{ij} = \sqrt{\Omega_i \Omega_j}$ is the geometric mean of individual beam solid angle, in which $\Omega_i = \int \dd^2 \hat{\vb*{n}} |A_i(\hat{\vb*{n}})|^2$, $N_\text{day}$ is the number of observation days.  $N_\text{red}$ is the number of redundant baseline, $t_\text{sid}$ is the sidereal day in second, $\Delta \nu$ is the bandwidth, $t_{\text{int}}$ is the single integration time, for the unpolarized case, $\delta_{ij}^{I} = \delta_{ij} / \sqrt{2}$. 
Below we assume all feeds have the same system temperature and primary beam profile.  We take the system temperatures as $90$ K for all feeds from the estimation in \cite{2020SCPMA..6329862L}, $t_{\text{int}} = 4$ s and $\Delta \nu = 122$ kHz for the current configuration of Tianlai cylinder pathfinder, and take $N_{\text{day}} = 1$ for a single sidereal day observation, or 30 as an imitation for multiple sidereal days stacking which increases the signal-to-noise ratio.

\subsection{Beam Coverage in Spherical Harmonics Space}
\label{sec:SHBC}

The angular resolution and sensitivity on the sky of a telescope is limited by its size and configuration. We take the maximum spherical harmonics degree $l$ and order $m$ that Tianlai cylinder pathfinder sensitive to at 750 MHz as
$$
l_{\rm max} = 2\pi D_{\rm max} / \lambda \approx 740, \qquad m_{\rm max} = 2\pi D_{\rm EW} / \lambda \approx 715,
$$
where $D_{\rm max}$ is the maximum dimension of the entire array, $D_{\rm EW}$ is the physical size in the East-West direction.

\begin{figure}
    \centering
    \includegraphics[width=0.6\textwidth]{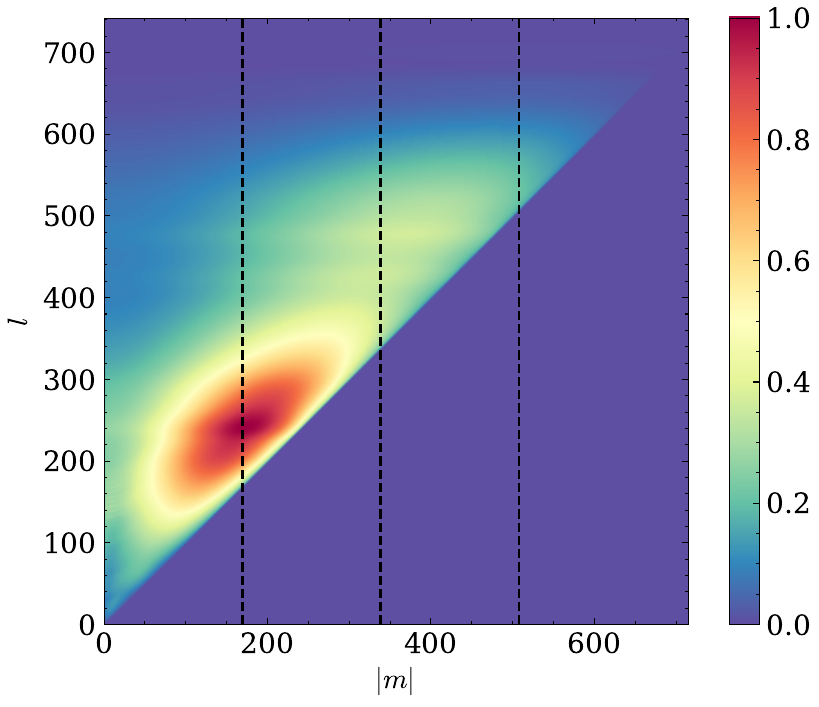}
    \caption{
    The spherical harmonic beam coverage $\mathbb{B}_{l|m|}$ of the Tianlai cylinder pathfinder array. The three vertical dotted lines indicate the $m$ values corresponding to the width of a single cylinder (15 m), the maximum baseline projected in the East-West direction (30 m), and the total width of the array (45 m).
    }
    \label{fig:tianlai-SHBC}
\end{figure}

We can define the spherical harmonic beam coverage of array by combining the beam transfer matrix from all baselines as
\begin{equation}
    \mathbb{B}_{l|m|} = \frac{\sum_{n=1}^{N} |B_{l|m|}^{n}|}{\mathrm{max}(\sum^{N}_{n=1} |B_{l|m|}^{n}|)},
    \label{eq:shbc}
\end{equation}
where $N$ is the total number of non-redundant baselines. This quantity can be used to describe the sensitivity of array on the sky in spherical harmonic space, analogy to the \textit{uv} coverage, which is used to define the spatial frequencies measured by the array.
In Fig. \ref{fig:tianlai-SHBC}, we show the beam coverage $\mathbb{B}_{l|m|}$ of the Tianlai cylinder pathfinder array. 

The resolution in the East-West direction of a baseline is determined by its projected distance along this direction. The limit on $m$ for a telescope is $m < 2\pi D_{\rm EW} \cos{\delta}/ \lambda$ by taking into account that the $m$ mode corresponds to the Fourier mode in the azimuthal direction, where $\delta$ is the latitude of the Tianlai site about $44^{\circ}$, thus $m \lesssim 510$ with $D_{\rm EW} = 45$, $\lambda \approx 0.3997$ at $750$ MHz. The beam coverage is expected to center at $(l,\; m) = (2\pi D / \lambda,\; 2\pi D_{\rm EW} \cos{\delta}/ \lambda)$. Owing to the massive isometric short baselines on the identical cylinder, we expect to see two areas in the spherical harmonic space where the Tianlai cylinder pathfinder having higher sensitivity, which can be estimated by
$$
(l_{\rm center},\; m_{\rm center}) \approx (2\pi \sqrt{D_{w}^{2} + b^2} / \lambda,\; 2\pi {D_{w}} \cos{\delta} / \lambda),
$$
with $D_{w}$ equals 1 or 2 times the cylinder width, and $b= 0.4$ m, these are at $(l,\; m) \approx (235,\; 170)$ and $(470,\; 340)$, which is indeed the case as can be seen in Fig. \ref{fig:tianlai-SHBC}.

Roughly, the sensitivity of our model telescope is relatively high in the range of about $[m,\; m+200]$ at $m \lesssim 400$, and decreases significantly at and outside the edge region (i.e., $m \sim 510$) that the telescope can reach. Consequently, when reconstructing the sky map from the spherical harmonic coefficient $a_{lm}$ in this paper, we filtered out all modes $l > 600$, as well as the $m = 0$ modes which measure the average over sidereal time.

\subsection{Linear Least-Squares Map-Maker}

The imaging procedures can be seen as trying to solve the solution of Equation \eqref{eq:mmode-3} for the spherical harmonic coefficients $\vb*{a}$ given the measurement $\vb*{v}$ and the beam matrix $\vb*{B}$. The least-squares solution to it, which minimizes $|| \vb*{v} - \vb*{Ba}||^2_2$, is given by
\begin{equation}
    \hat{\vb*{a}}_{\mathrm{LS}} = (\vb*{B}^{*}\vb*{B})^{-1} \vb*{B}^{*} \vb*{v},
    \label{eq:ls-solver}
\end{equation}
where $*$ denotes the conjugate transpose. 

In our current work, before making the general linear least-squares solution, we first subtract the contributions of four very strong radio sources to the visibilities, i.e. Cas A, Cyg A, Tau A, and Vir A, which can be treated as point sources at known locations for the current Tianlai cylinder array pathfinder. This reduces the dynamic range of the input map. If they are not subtracted, the reconstructed map will show some apparent sidelobes around their positions. 

\begin{figure}
    \centering
    \includegraphics[width=0.6\textwidth]{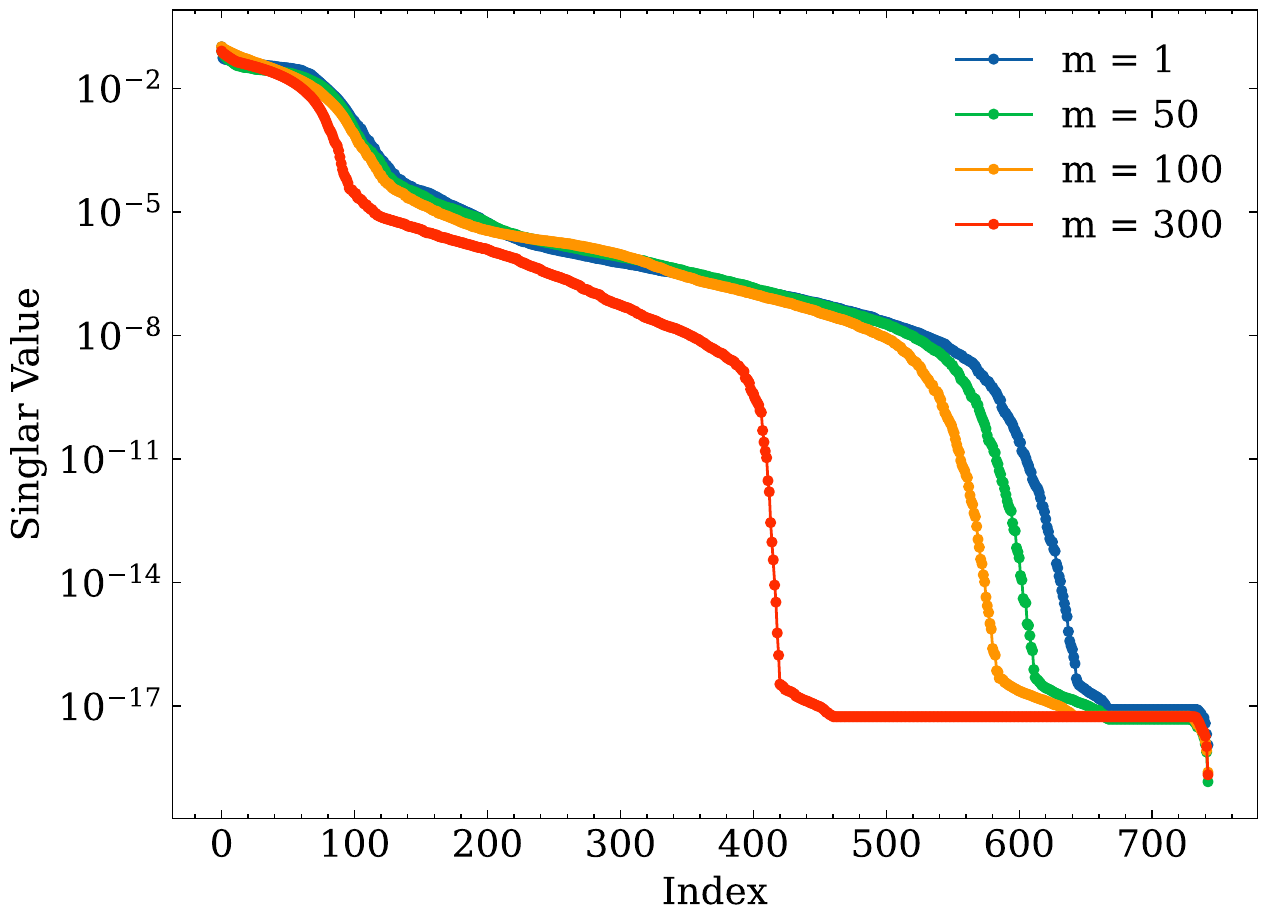}
    \caption{The singular values of matrix $\vb*{B}$ for several $m$ values.}
    \label{fig:B_sv}
\end{figure}

The matrix $\vb*{B}$ usually does not have full rank, as there are unmeasured modes due to incomplete $uv$-coverage and incomplete coverage of the full sky. In Fig. \ref{fig:B_sv}, we plot the singular values of several $\vb*{B}$, a cluster of small singular values approaching zero results in these matrices being rank-deficient. Hence $\vb*{B}^{*}\vb*{B}$ is not invertible, which makes the problem ill-posed. 
The solution to ill-posed inverse problems is usually unstable, which may deviate greatly from the true one in the presence of noise.

\section{Regularization}
\label{sec:reg-solver}
A common approach to ill-posed problems is regularization (see, e.g. \cite{engl1996regularization, hansen1998rank}), which gives a stable approximate solution by imposing additional constraints on it. 
The common regularization methods include Truncated Singular Value Decomposition (TSVD) and Tikhonov regularization. 

\subsection{Truncated Singular Value Decomposition}

For the least-squares solution (Equation \eqref{eq:ls-solver}), the matrix $\vb*{B}$ generally does not have full rank, $\vb*{B}^{*}\vb*{B}$ is then singular and not invertible.
With the singular values decomposition of the $m \times n$ ($m \ge n$) matrix $\vb*{B}$,
$$
\vb*{B} = \vb*{U \Sigma V^*} = \sum^{n}_{i=1} \vb*{u}_i \sigma_i \vb*{v}_i^{*},
$$
where $\vb*{U} = (\vb*{u}_1, \vb*{u}_2, \cdots, \vb*{u}_n)$ and $\vb*{V} = (\vb*{v}_1, \vb*{v}_2, \cdots, \vb*{v}_n)$ are unitary matrices, and the diagonal matrix $\vb*{\Sigma} = \mathrm{diag}(\sigma_1, \sigma_2, \cdots, \sigma_n)$ has non-negative elements in non-increasing order.
The Moore-Penrose pseudo-inverse of matrix $\vb*{B}$ is defined as
\begin{equation*}
    \vb*{B}^{+} = \vb*{V} \vb*{\Sigma}^{-1} \vb*{U}^{*},
\end{equation*}
where $+$ denotes the pseudo-inverse, the diagonal element of matrix $\vb*{\Sigma}^{-1}$ is the reciprocal of each non-zero element on the diagonal of $\vb*{\Sigma}$, leaving the zeros in place. The minimum-norm least-squares solution to Equation \eqref{eq:mmode-3} in terms of the Moore-Penrose pseudo-inverse is then given by
\begin{align}
    \nonumber
    \hat{\vb*{a}}_{\rm LS} &= \vb*{B}^{+}\vb*{v} = \vb*{V} \vb*{\Sigma}^{-1} \vb*{U}^{*} \vb*{v} \\
    &= \sum^{r}_{i=1} \frac{\vb*{u}^{*}_{i} \vb*{v}}{\sigma_{i}} \vb*{v}_i,
    \label{eq:ls-svd}
\end{align}
where $r = {\rm rank}(\vb*{B})$.
For the noisy measurement $\vb*{v} = \vb*{v}_{0} + \vb*{n}$, where $\vb*{v}_{0} = \vb*{B}\vb*{a}_{\rm true}$, this solution becomes
\begin{equation}
    \hat{\vb*{a}}_{\mathrm{LS}} = \sum^{r}_{i=1} \frac{\vb*{u}^{*}_{i} \vb*{v}_0}{\sigma_{i}} \vb*{v}_i + \sum^{r}_{i=1} \frac{\vb*{u}^{*}_{i} \vb*{n}}{\sigma_{i}} \vb*{v}_i ,
\end{equation}
the first term gives the true solution, while the second term gives the contribution from noise. The latter might be magnified by those extremely small singular values $\sigma_{i}$ of $\vb*{B}$, making the solution unstable and heavily contaminated by noise.

Regularization is a technique to solve the problem above by dampening or filtering out those unwanted components corresponding to the small singular values, leading to an approximate but stable solution. In terms of the singular values decomposition of matrix $\vb*{B}$, the regularized solution to Equation \eqref{eq:mmode-3} can be written as
\begin{equation}
    \hat{\vb*{a}}_{\mathrm{reg}} = \sum^{r}_{i=1} f_i \frac{\vb*{u}^{*}_{i} \vb*{v}}{\sigma_{i}} \vb*{v}_i,
    \label{eq:reg_sol_svd}
\end{equation}
where $f_i$ are the \textit{filter factors}. Different regularization methods can be defined by choosing suitable filter factors.

\begin{figure}
    \centering
    \includegraphics[width=0.33\textwidth]{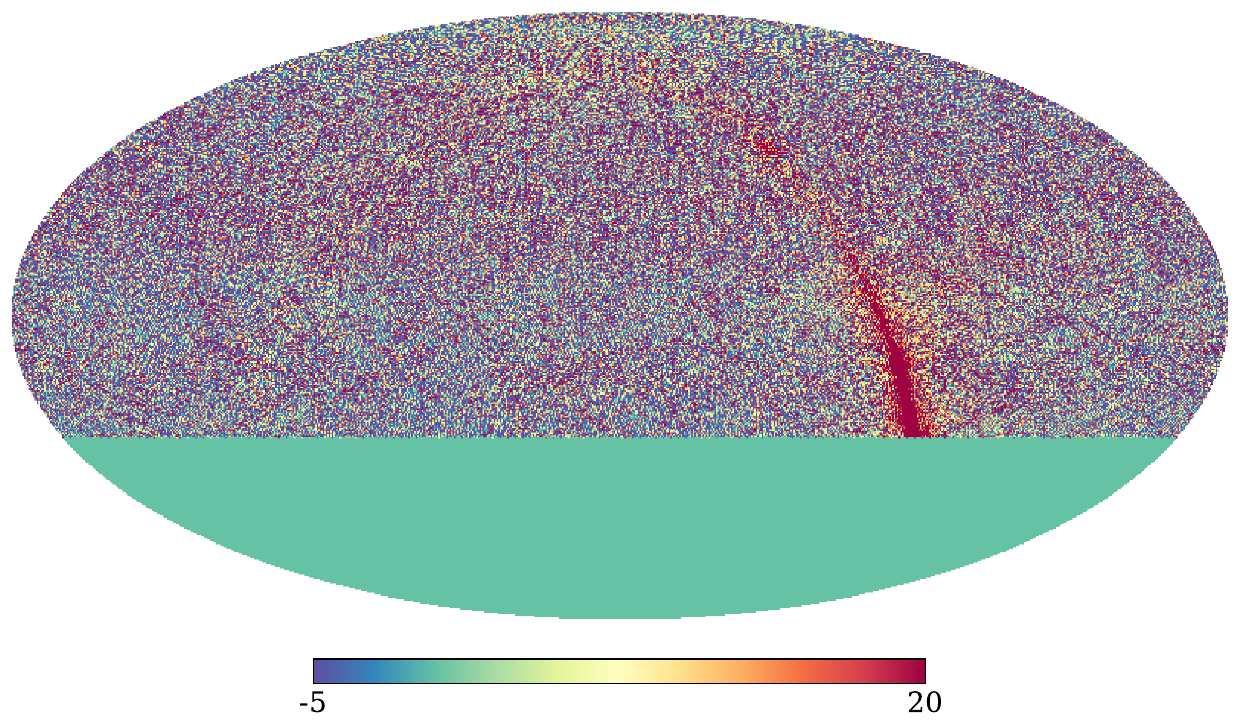}
    \includegraphics[width=0.33\textwidth]{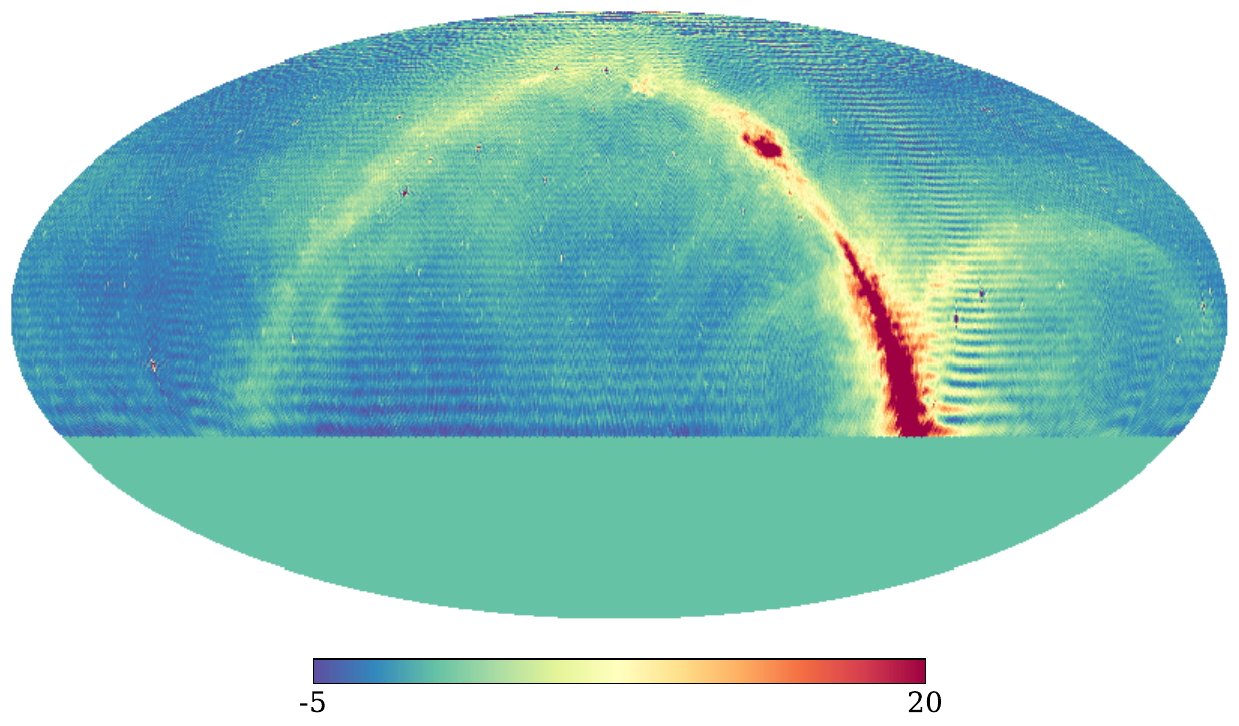} 
    \includegraphics[width=0.33\textwidth]{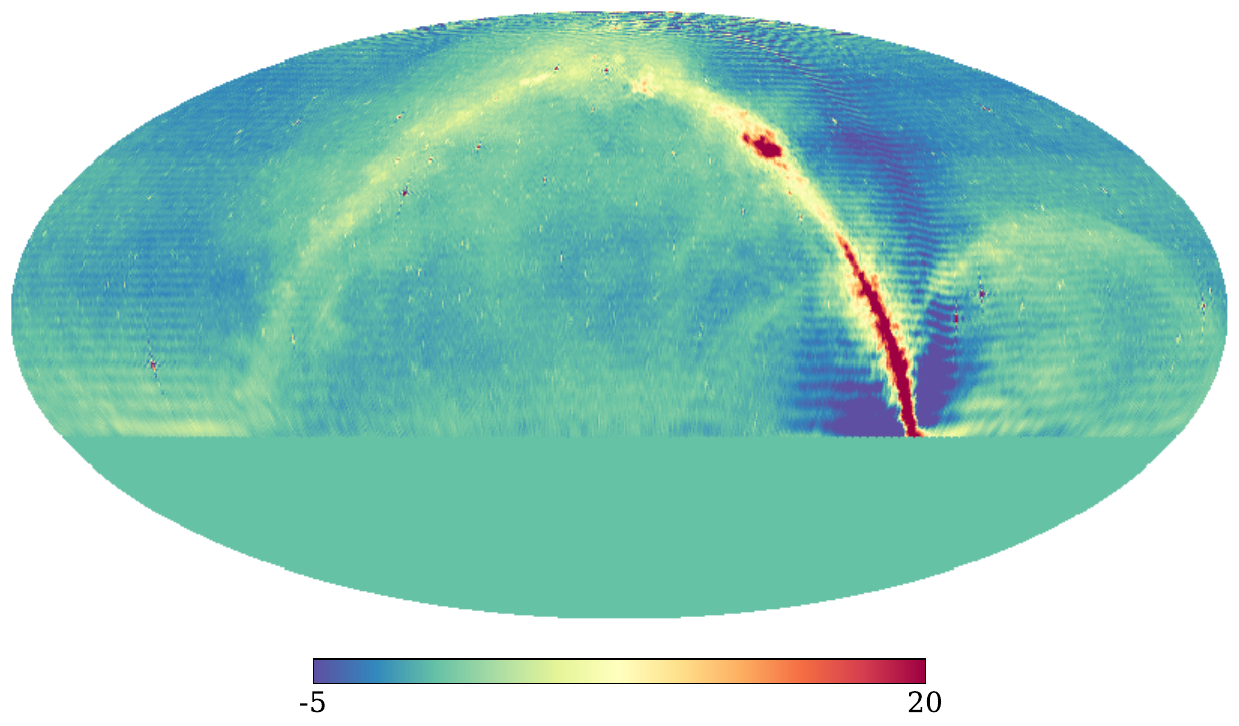}
    \caption{The reconstructed sky maps with different truncation thresholds for the data with a noise level corresponding to 30-day integration time. From left to right, $\epsilon=1\times 10^{-4}, 1\times10^{-3}$, and $5\times 10^{-3}$. The four brightest point sources (i.e. Cas A, Cyg A, Tau A and Vir A) have been removed in the visibility before reconstruction.}
    \label{fig:re-map-tsvd-diff-reg-par}
\end{figure}

For the TSVD regularization, the filter factors are
\begin{eqnarray}
    f_{i}^{\text{TSVD}} =
    \begin{cases}
    1 & \sigma_i \geq \sigma_k \\
    0 & \sigma_i < \sigma_k
    \end{cases},
    \label{eq:tsvd-filter-factor}
\end{eqnarray}
the pseudo-inverse $\vb*{B}^{+}$ in Equation \eqref{eq:ls-svd} is replaced by its rank-$k$ approximation, with the remaining components  filtered out. In paper I, we used the TSVD regularization to compute the regularized solution of Equation \eqref{eq:mmode-3}, where the truncation thresholds $\epsilon$ satisfies $\sigma_{k+1} < \epsilon \times {\rm max}(\sigma_i) \leq \sigma_{k}$, and we selected $\epsilon = 1 \times 10^{-3}$ as the default.

As an illustration, we show in Fig.~\ref{fig:re-map-tsvd-diff-reg-par} the reconstructed maps with three truncation thresholds $\epsilon$, which can represent the case with a too small, moderate and too large $\epsilon$, respectively.
Noise in the reconstructed map is significantly amplified if there is no regularization or with a too small threshold value $\epsilon$, the image is totally dominated by noise. For a too large $\epsilon$, although noise is not significantly amplified, the reconstructed signal may fail to capture the true signal completely in certain modes, so again the resulting image is not optimal. For a properly chosen regularization value, we can see the reconstruction result is relatively good, with the input sky map well recovered. However, as we noted in paper I, even in this case, near the bright part of the Galactic plane, there are comb-like artifacts, which are produced by the incomplete reconstruction of certain modes which are truncated, especially those with small $m$ values in our case.

\subsection{Tikhonov Regularization}

Another widely used regularization method is Tikhonov regularization. In contrast to TSVD regularization, which imposes a hard threshold for the components corresponding to those small singular values, Tikhonov regularization aims to suppress undesirable modes by solving for $\vb*{a}$ that minimizes
\begin{equation}
    \underset{\vb*{a}}{\text{argmin}} \{\norm{\vb*{v} - \vb*{Ba}}_2^2 + \lambda^2 \norm{\vb*{L(a - a_0)}}_2^2\},
\label{eq:tik_general}
\end{equation}
where $|| \cdot ||_2$ denotes the Frobenius norm, $\lambda$ is the regularization parameter, $\vb*{a}_0$ is a prior of $\vb*{a}$, which can be set to $\vb*{a}_0 = 0$ if prior information is unavailable, and $\vb*{L}$ is called the Tikhonov regularization matrix. There are several common choices for $\vb*{L}$, including the identity matrix or a discrete approximation of the derivative operator.
This minimization problem gives solution
\begin{equation}
    \hat{\vb*{a}} = (\vb*{B}^{*}\vb*{B} + \lambda^2 \vb*{L}^{*} \vb*{L})^{-1} (\vb*{B}^{*} \vb*{v} +  \lambda^2 \vb*{L}^{*} \vb*{L} \vb*{a}_0).
    \label{eq:tk-reg-1}
\end{equation}

In a simple case where $\vb*{L} = \vb*{I}$, the Tikhonov regularization is referred to as the \textit{standard form}.
The more general form of Tikhonov regularization, where $\vb*{L} \neq \vb*{I}$ and prior information about $\vb*{a}$ and $\vb*{n}$ is available, can be transformed into the standard form \citep{elden1977algorithms}. For simplicity, we adopt $\vb*{L} = \vb*{I}$ and $\vb*{a}_0 = 0$, then the objective function becomes
\begin{equation}
    \underset{\vb*{a}}{\text{argmin}} \{\norm{\vb*{v} - \vb*{Ba}}_2^2 + \lambda^2 \norm{\vb*{a}}_2^2\},
\end{equation}
and Equation \eqref{eq:tk-reg-1} is reduced to
\begin{equation}
    \hat{\vb*{a}}_{\rm tk} = (\vb*{B}^{*}\vb*{B} + \lambda^2 \vb*{I})^{-1} \vb*{B}^{*} \vb*{v}.
    \label{eq:tk-reg-2}
\end{equation}
It can be expressed in the form of Equation \eqref{eq:reg_sol_svd} with filter factors as
\begin{eqnarray}
    f_{i}^{\text{tk}} = \frac{\sigma_{i}^{2}}{\sigma_{i}^{2} + \lambda^2} \eqsim
    \begin{cases}
        1 & \sigma_{i} \gg \lambda \\
        0 & \sigma_{i} \ll \lambda
    \end{cases}.
\end{eqnarray}

Unlike the TSVD regularization which simply trims out the modes corresponding to small eigenvalues, the Tikhonov regularization makes a gradual and smooth suppression to all modes. The degree of smoothness can be adjusted by the choice of the regularization parameter $\lambda$, a small regularization parameter value yields a solution which is closer to the one obtained from TSVD regularization.

\begin{figure}
    \centering
    \includegraphics[width=0.33\textwidth]{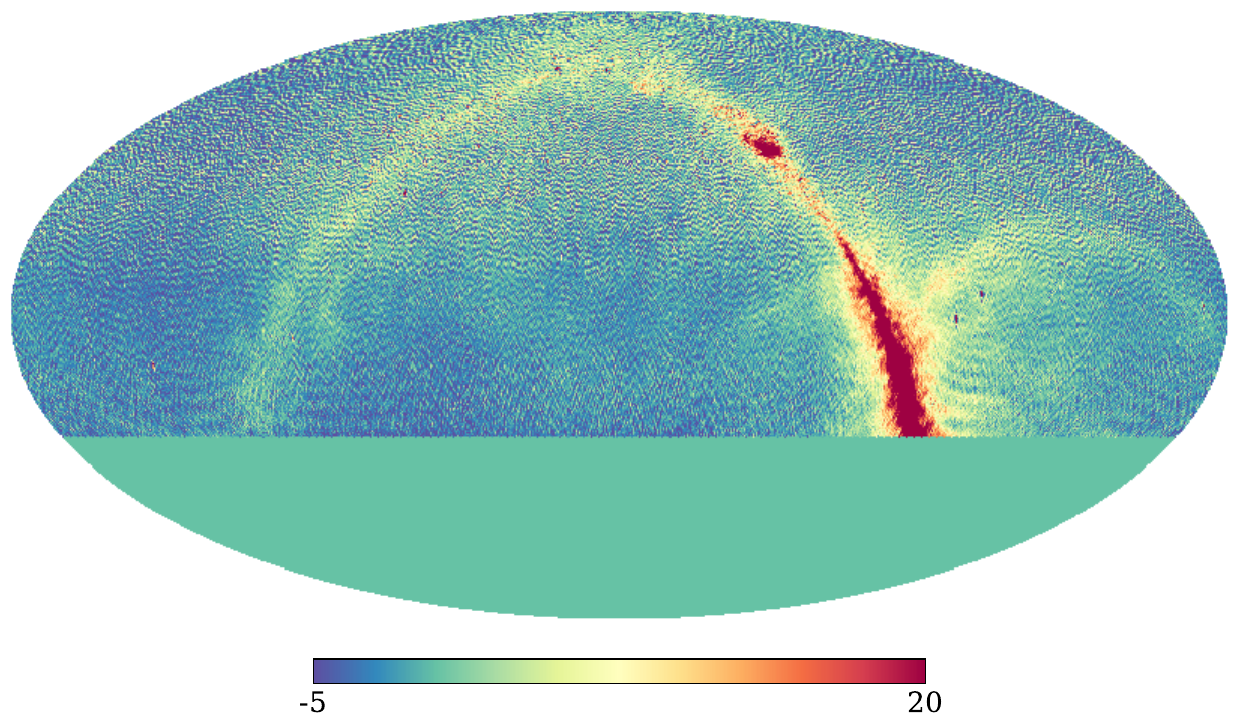}  \includegraphics[width=0.33\textwidth]{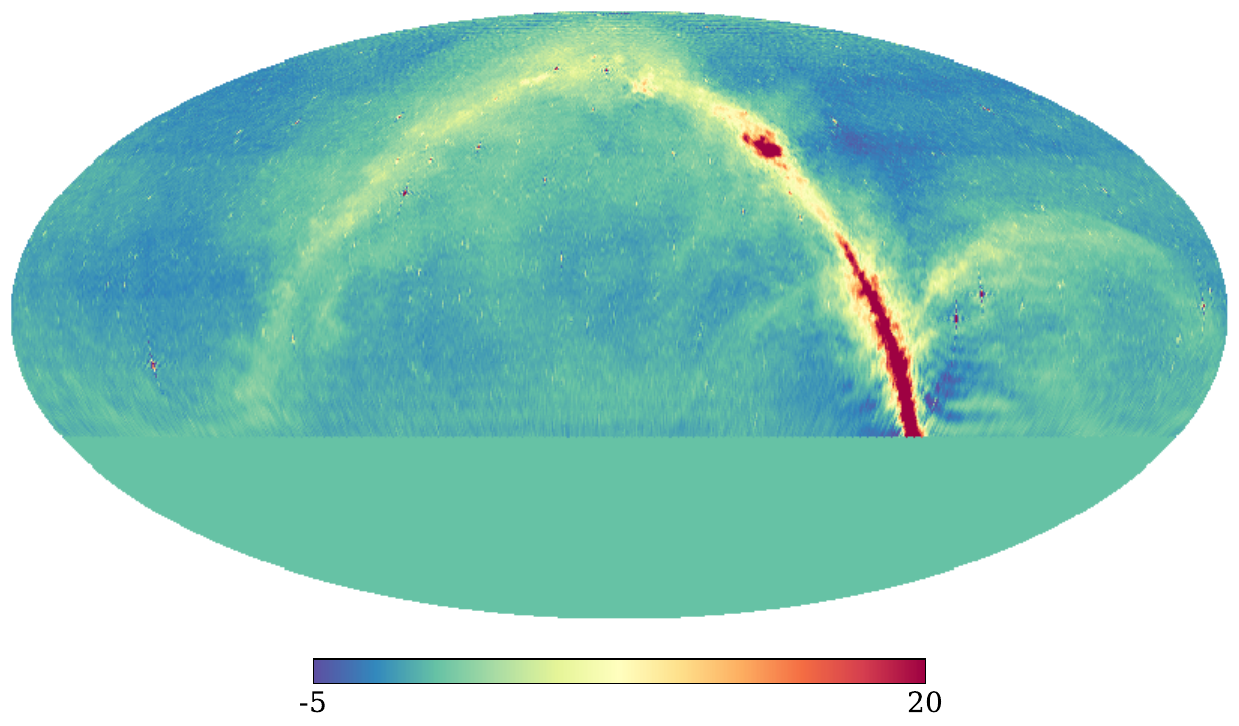}  \includegraphics[width=0.33\textwidth]{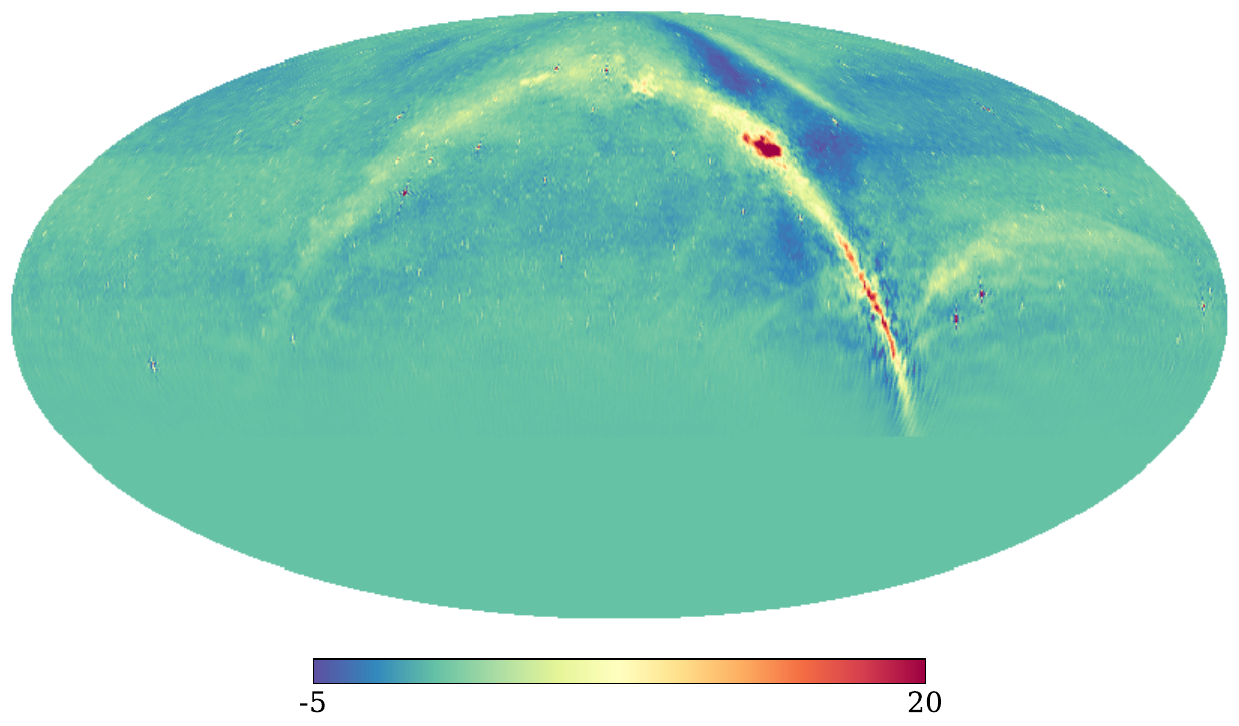}
    \caption{The reconstructed sky map with different Tikhonov regularization parameter $\lambda$. From left to right, $\lambda^2 =$ $10^{-9}$, $9 \times 10^{-8}$, and $10^{-5}$.}
    \label{fig:re-map-tk-diff-reg-par}
\end{figure}

In Fig. \ref{fig:re-map-tk-diff-reg-par}, we illustrate the Tikhonov regularized sky maps with different regularization parameter $\lambda$. For a small $\lambda$, as shown in the left panel, the noise in the regularized map is amplified obviously. The noise becomes less pronounced with a moderate $\lambda$ value, as shown in the center panel. The comb-like artifact around the Galactic Center shown in Fig. \ref{fig:re-map-tsvd-diff-reg-par} is also less prominent, perhaps thanks to the gradual suppression on the modes in the Tikhonov regularization. However, for the case with an excessively large $\lambda$ shown in the right panel as an example, the underlying signal is also suppressed, generating a more bland map than the actual sky. Furthermore, accompanied with the diminished actual sky signal, the sidelobes from some bright sources are more pronounced, e.g. an arc structure becomes noticeable near the north celestial pole and above the Cyg A.
An appropriate selection of $\lambda$ value, which strikes a balance between noise amplification and information loss, is expected to yield a satisfactory map that contains moderate noise while faithfully preserving the true sky as much as possible.

The regularized solution error is given by
\begin{equation}
    \vb*{a}_{\rm true} - \vb*{a}_{\rm reg} = \sum^{n}_{i=1} (1 - f_i) \vb*{v}_i^{*} (\vb*{a}_{\rm true}) \vb*{v}_i - \sum^{n}_{i=1} f_i \frac{\vb*{u}_i^{*} \vb*{n}}{\sigma_i} \vb*{v}_i,
    \label{eq:reg_solver_error}
\end{equation}
where the first term corresponds to the \textit{regularization error}, arising from the regularization of true signal $\vb*{a}_{\rm true}$, and the second term represents the \textit{perturbation error}, attributed to the presence of noise. The filter factors $f_i$ can be $f_i^{\rm tsvd}$ or $f_i^{\rm tk}$. For Tikhonov regularization, as $\lambda \rightarrow 0$, the filter factors $f_{i}^{\rm tk} \rightarrow 1$, the regularization error approaches zero, but the perturbation error might be large, the solution tends to be noisy. As $\lambda$ increases, the filter factors $f_{i}^{\rm tk} \rightarrow 0$, leading to a smaller perturbation error but a larger regularization error, and the regularized solution is oversmoothed and approaches zero.
The key to obtaining a good regularized solution is to choose an optimal regularization parameter $\lambda$ in Tikhonov regularization or $k$ in TSVD regularization to balance the two errors.

\subsection{The Choice of Regularization Parameter}

Although choosing the optimal regularization parameter that balances the trade-off between the noise amplification and signal recovery is not always easy, which depends on particular problems including factors such as the specific array configuration, and the level of noise in the data. There are still several approaches available for choosing the regularization parameter that near the optimal one. Below we adopt the notation used in Tikhonov regularization (i.e. take $\lambda$ as regularization parameter) for our description, these methods are also applicable to TSVD regularization.
If the statistics of the noise is well known, $\lambda$ can be chosen by applying the discrepancy principle \citep{engl_discrepancy_1987}, such that the residue is at a comparable level to that of the noise, $||\vb*{B}\vb*{a}_{\lambda} - \vb*{v}||_2 \leq \eta ||\vb*{n}||_2$, where $\vb*{a}_{\lambda}$ is the solution given by Equation \eqref{eq:tk-reg-2} with the regularization parameter value $\lambda$, and $\eta$ is a user-specified constant to constrain the bound. 
If the error is unknown, techniques such as the generalized cross validation \citep{golub1979generalized} or the L-curve criterion \citep{Hansen1993TheUO} are usually applied to search for the appropriate regularization parameter.

\paragraph{\textbf{Generalized Cross Validation (GCV)}} The GCV method \citep{golub1979generalized} is based on the following idea: for a linear vector equation $\vb*{Ax} = \vb*{b}$, if we drop a data point $\vb*{b}^i$ from the vector $\vb*{b}$, and obtain a regularized solution $\vb*{x}_{\lambda}^{i}$ from the rest of vector components $\vb*{A}([1:i-1,\; i+1:m], \; :) \vb*{x}_{\lambda}^{i} \approx \vb*{b}([1:i-1,\; i+1:m])$, then the value $\vb*{A}(i,\; :) \vb*{x}_{\lambda}^{i}$ should be close to the excluded value $\vb*{b}^{i}$ if a reasonable parameter $\lambda$ is chosen (c.f. Chapter 4 of \citealt{wahba1990spline}). This results in the selection of a parameter $\lambda$ that minimizes the GCV function
\begin{equation}
    \mathcal{G}(\lambda) = \frac{||\vb*{B} \vb*{a}_{\lambda} - \vb*{v}||_{2}^{2}}{(\text{trace}(\vb*{I} - \vb*{BB}^{\#}(\lambda)))^2},
    \label{eq:gcv-function}
\end{equation}
where $\vb*{B}^{\#}(\lambda) \equiv (\vb*{B}^{*} \vb*{B} + \lambda^2 \vb*{I})^{-1} \vb*{B}^{*}$ is referred to as the \textit{regularized inverse}, $\vb*{a}_{\lambda} = \vb*{B}^{\#} \vb*{v}$ is the corresponding regularized solution.

\begin{figure}
    \centering
    \includegraphics[width=0.5\textwidth]{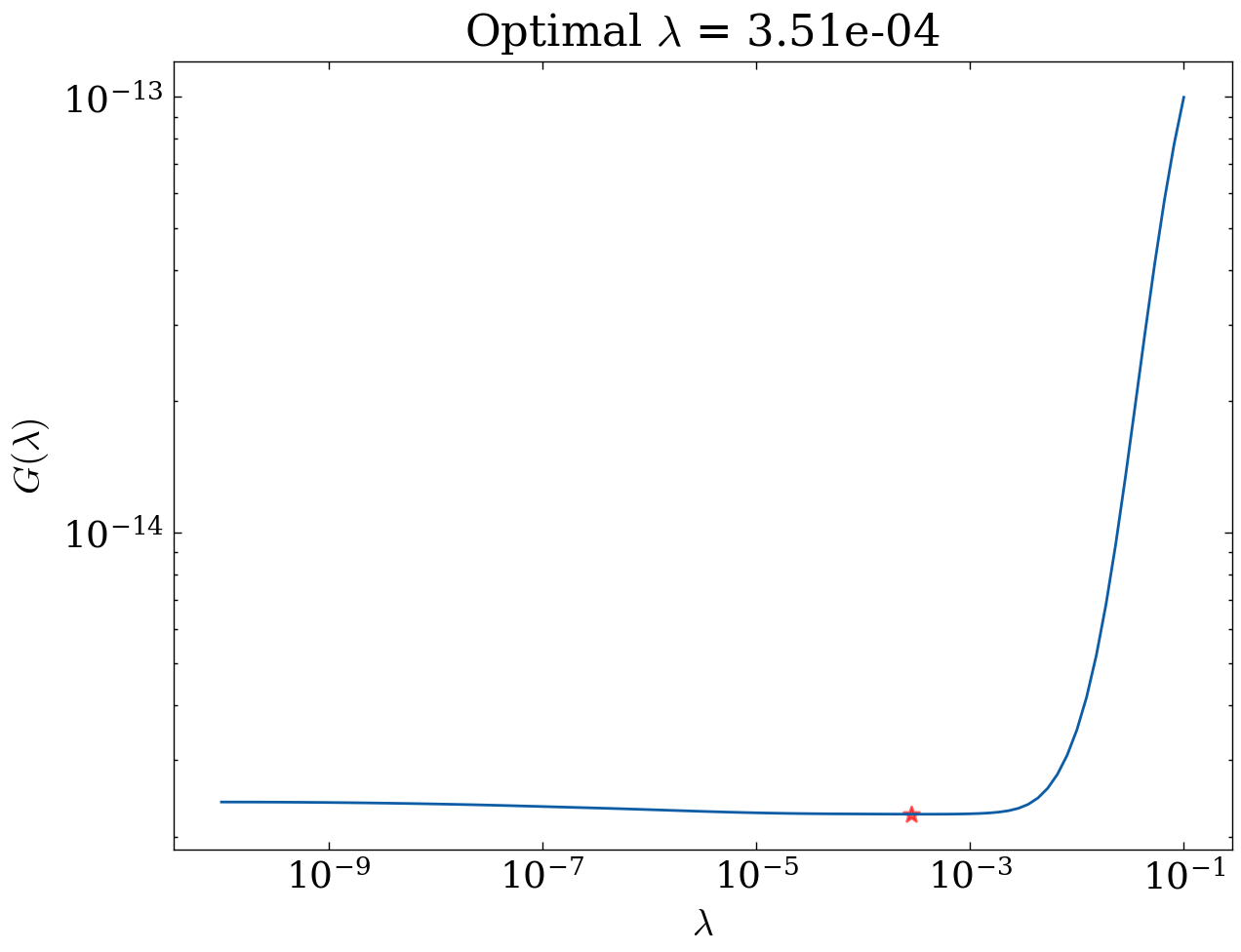}
    \caption{The GCV function as a function of $\lambda$ for $m = 200$ case. The red star marks the minimum point of the GCV function, the corresponding $\lambda$ gives the optimal regularization parameter by this method.}
    \label{fig:gcv}
\end{figure}

\begin{figure}
    \centering
    \includegraphics[width=0.9\textwidth]{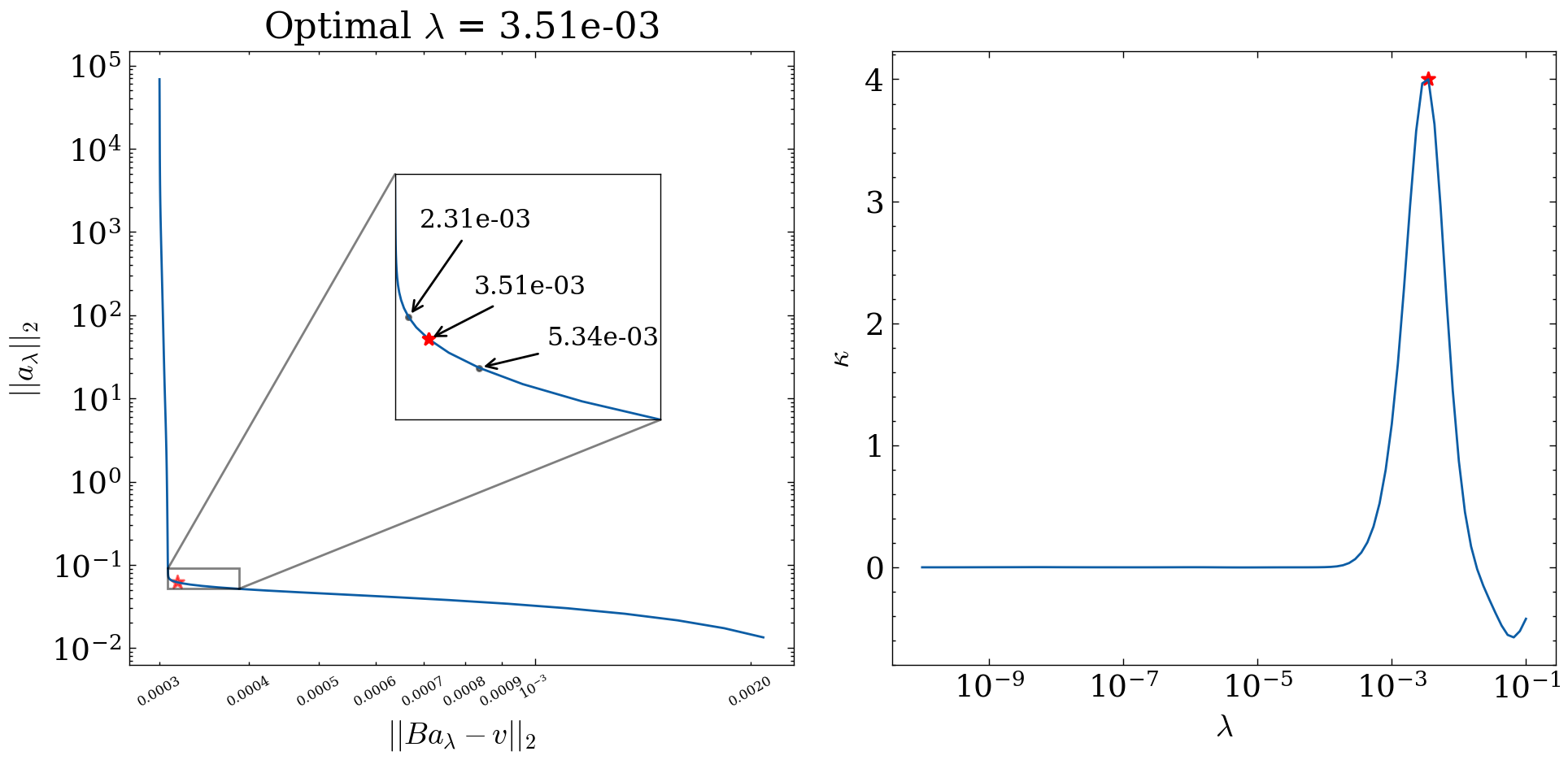}
    \caption{The L-curve (left) and the corresponding curvature (right) $\kappa$ as a function of $\lambda$ for $m = 200$ case. The red star marks the point of maximum curvature at the corner of the L-curve.}
    \label{fig:l-curve}
\end{figure}

For illustration, in Fig. \ref{fig:gcv} we show the logarithmic plot of the calculated GCV function applied to the data with 30-day integration noise for the case of $m = 200$. In our computation, the $\lambda$ values vary from $10^{-10}$ to $10^{-1}$ and are sampled evenly on logarithmic scale. The GCV function decreases slowly as $\lambda$ increases, until at some point increases rapidly, and the minimum is reached before this rapid rise, which is marked by a red star in the figure.

\paragraph{\textbf{L-curve}} In a \textit{log-log} scale plot depicting the residual norm $||\vb*{B}\vb*{a}_{\lambda} - \vb*{v}||_2$ versus the solution norm $||\vb*{a}_{\lambda}||_{2}$, the resulting curve typically exhibits a L-shaped profile, which is comprised of a flat part where regularization error dominates at large $\lambda$, and a steep part where the perturbation error dominates at small $\lambda$. The optimal $\lambda$ value should be chosen to balance these two errors, which corresponds to the corner of the L-curve, and can be identified by locating the maximum curvature of the curve \citep{hansen1999curve}.
Let $\hat{\eta} = \log ||\vb*{a}_{\lambda}||_2^{2}$, $\hat{\rho} = \log ||\vb*{B}\vb*{a}_{\lambda} - \vb*{v}||_2^2$, then the curvature of the L-curve is given by
\begin{equation}
    \kappa = 2 \frac{\hat{\rho}' \hat{\eta}'' - \hat{\rho}'' \hat{\eta}'}{((\hat{\rho}')^2 + (\hat{\eta}')^2)^{3/2}}.
    \label{eq:l-curve-curvature}
\end{equation}
In Fig.\ref{fig:l-curve} we plot the L-curve (left) and the corresponding curvature $\kappa$ (right) applied to data with the 30-day integration noise for the $m = 200$ case, the red star marks the optimal parameter $\lambda$ determined using this method.

\begin{figure}
    \centering
    \includegraphics[width=0.45\textwidth]{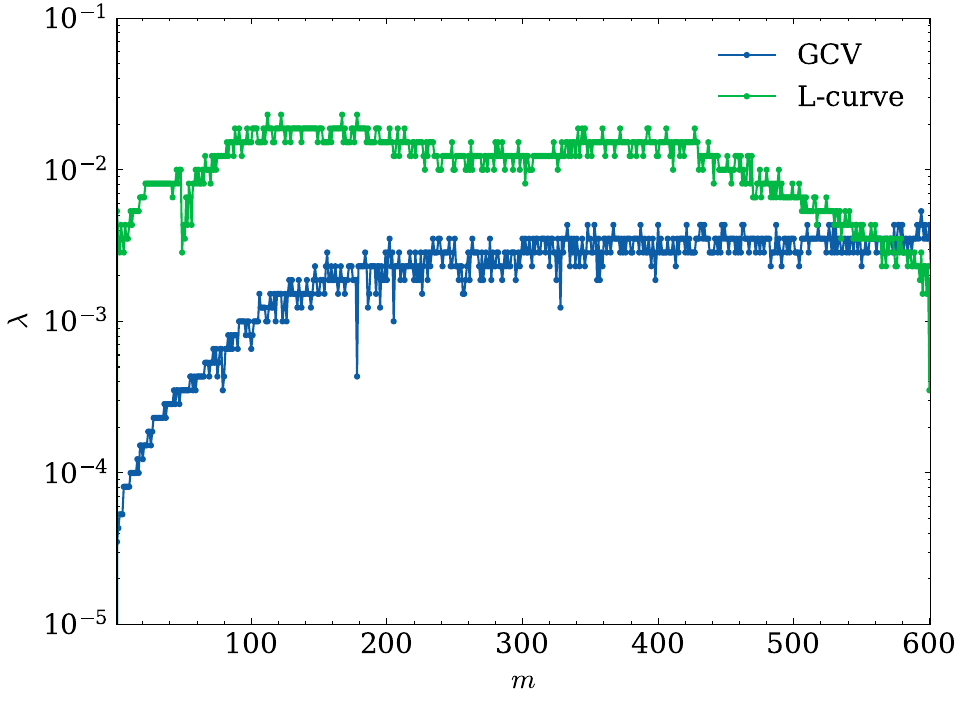}
    \includegraphics[width=0.45\textwidth]{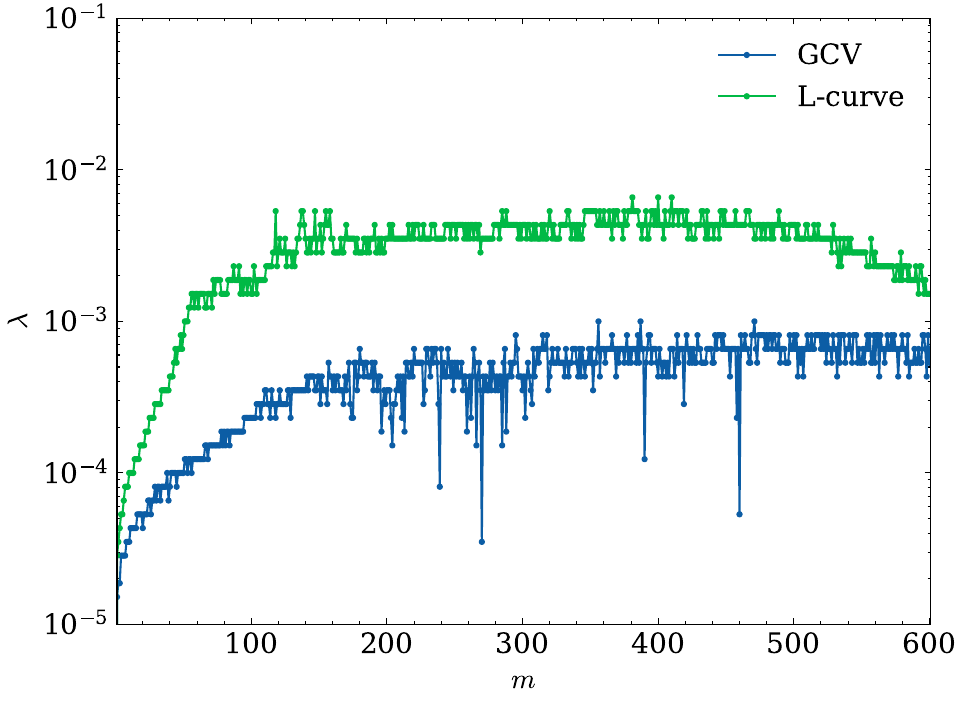}
    \caption{The optimal regularization parameters for all $0 < m \leq 600$ with GCV and L-curve method considering 1-day noise level (left) and 30-day noise level (right).}
    \label{fig:lamb_opt_comp}
\end{figure}

The optimal values for the regularization parameter $\lambda$ as determined by the GCV criterion and the L-curve criterion for all $m$ cases are shown in Fig. \ref{fig:lamb_opt_comp}.
Comparing the optimal values determined by these two criteria, it is observed that the optimal $\lambda$ obtained from the L-curve criterion is typically larger than the one provided by the GCV criterion at the same $m$, especially for the case of higher noise (e.g., 1-day integration noise level). The optimal $\lambda$ values increase as $m$ at $m \lesssim 100$, and then are relatively stable for the rest $m$. 

\begin{figure}
    \centering
    \includegraphics[width=0.45\textwidth]{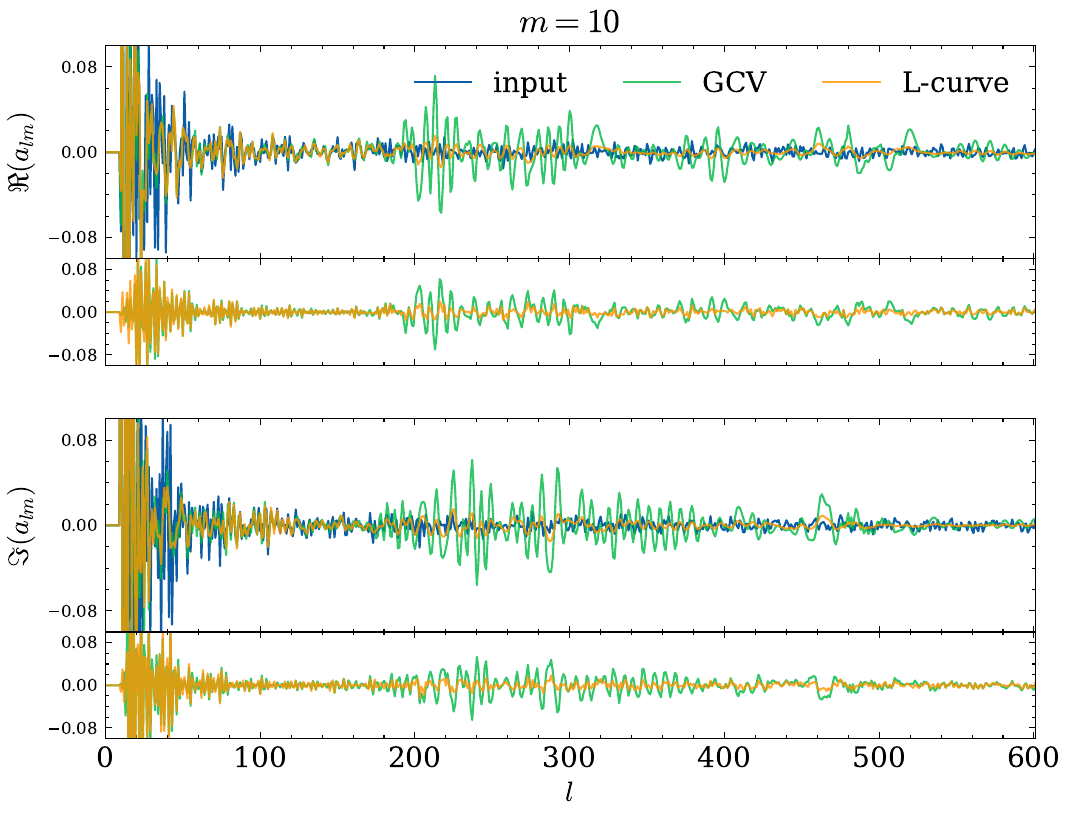}
    \includegraphics[width=0.45\textwidth]{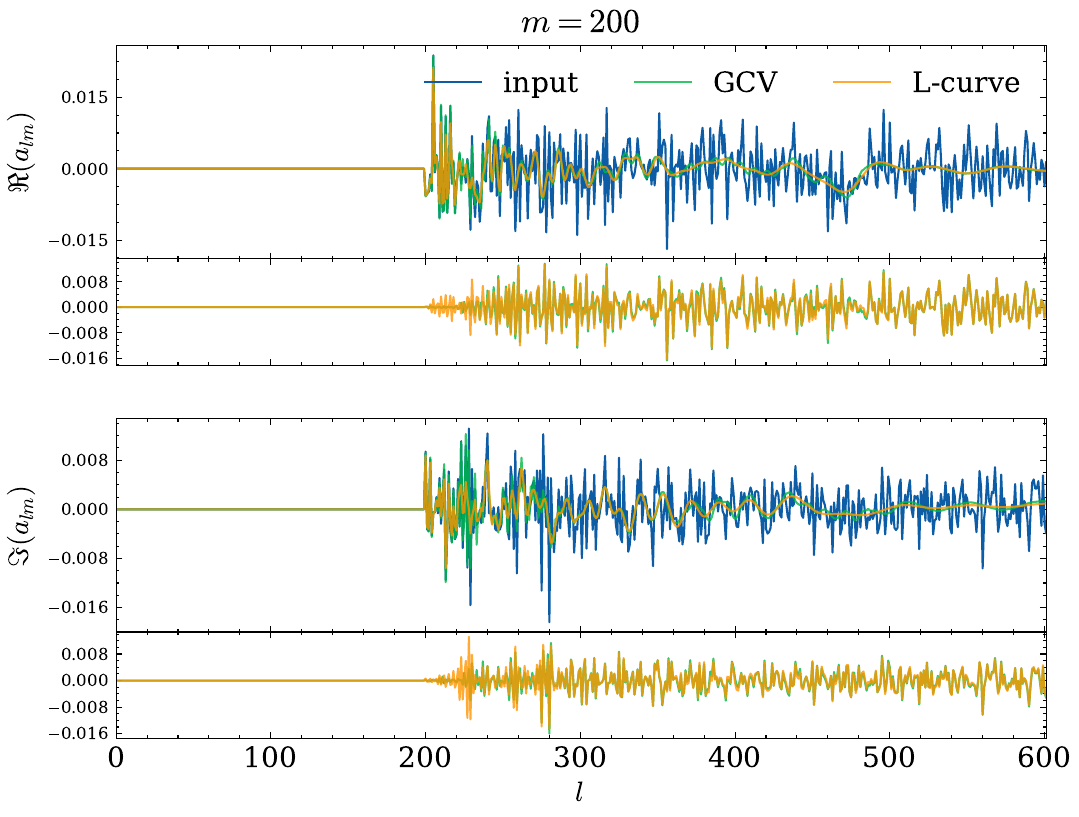}
    \caption{The comparison between the input (blue) and the solved spherical harmonics coefficients obtained using the Tikhonov regularization, with the regularization parameter determined by the GCV (green) and L-curve (orange) methods, for the $m = 10$ (left) and $m = 200$ (right) mode, in the case of 30-day noise level. The top sub-figure of each column shows the real part, and the bottom one gives the imaginary part. The bottom panel of each sub-figure shows the residue between the input and the solution.
    }
    \label{fig:reg-re-alm-1day}
\end{figure}

In Fig.~\ref{fig:reg-re-alm-1day}, we present a comparison between the input $a_{lm}$ and the Tikhonov regularized solution to the data with 30-day integration noise for the $m = 10$ (left) and $m = 200$ (right) cases as an example, where the regularization parameter $\lambda$ are given by the two criteria.
The real and imaginary parts are plotted in the top and bottom sub-figures, and the residues are plotted in the bottom panel of each sub-figure.
As expected, a smaller regularization parameter obtained from the GCV criterion results in a solution that is slightly closer to the true value, especially at the smaller $l$s,  but it may also amplify the noise for certain modes, as illustrated here in the region $l \in [200, \; 400]$ for the $m = 10$ case. While for the $m = 200$ case as shown, with the smaller regularization parameter provided by GCV criterion, the residue is smaller without the cost of amplifying the noise.

\begin{figure}
    \centering
    \includegraphics[width=0.45\textwidth]{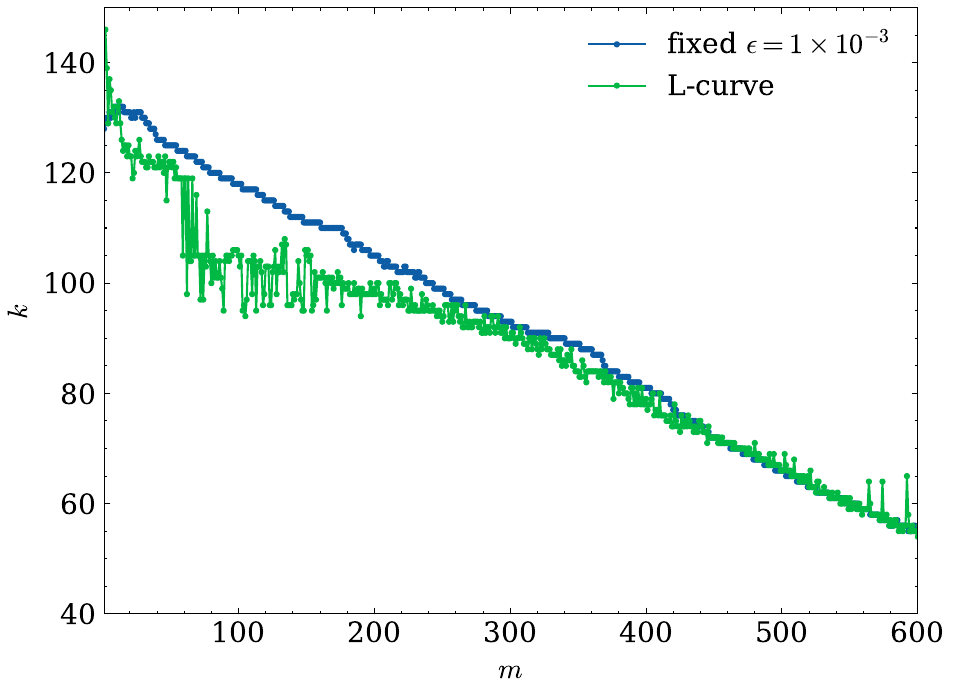}
    \caption{The truncation parameter $k$ for TSVD regularization obtained by the L-curve criteria for data with the 30-day integration noise. For comparison, the case for fixed $\epsilon=10^{-3}$ is also plotted. }
    \label{fig:tsvd-lcurve-ndays-30}
\end{figure}

Moreover, we apply the L-curve criterion to choose the truncation threshold value $\epsilon$ or $k$ in Equation \eqref{eq:tsvd-filter-factor} for the TSVD regularization. we sample 100 values of $\epsilon$ evenly on a logarithmic scale ranging from $1 \times 10^{-4}$ to $1 \times 10^{-1}$ to compute the points on the TSVD L-curve $(||\vb*{B}\vb*{a}_{k} - \vb*{v}||_2, ||\vb*{a}_{k}||_2)$. Different from the case of Tikhonov regularization, the points on TSVD regularization L-curve are discrete, a quadratic spline interpolation is applied to these discrete values to compute the curvature of the L-curve and select the corresponding sampled $\epsilon$ value closest to the maximum curvature. In Fig. \ref{fig:tsvd-lcurve-ndays-30}, we show result for the data with 30-day integration noise. For comparison, the case for fixed $\epsilon=1\times 10^{-3}$ is also plotted. We can see that at large $m$ the two curves almost coincide, but at some relative smaller $m$ the L-curve suggests a smaller $k$, indicating that more low sensitive modes are trimmed off to avoid amplifying noise.

\begin{figure}
    \centering
    \includegraphics[width=0.33\textwidth]{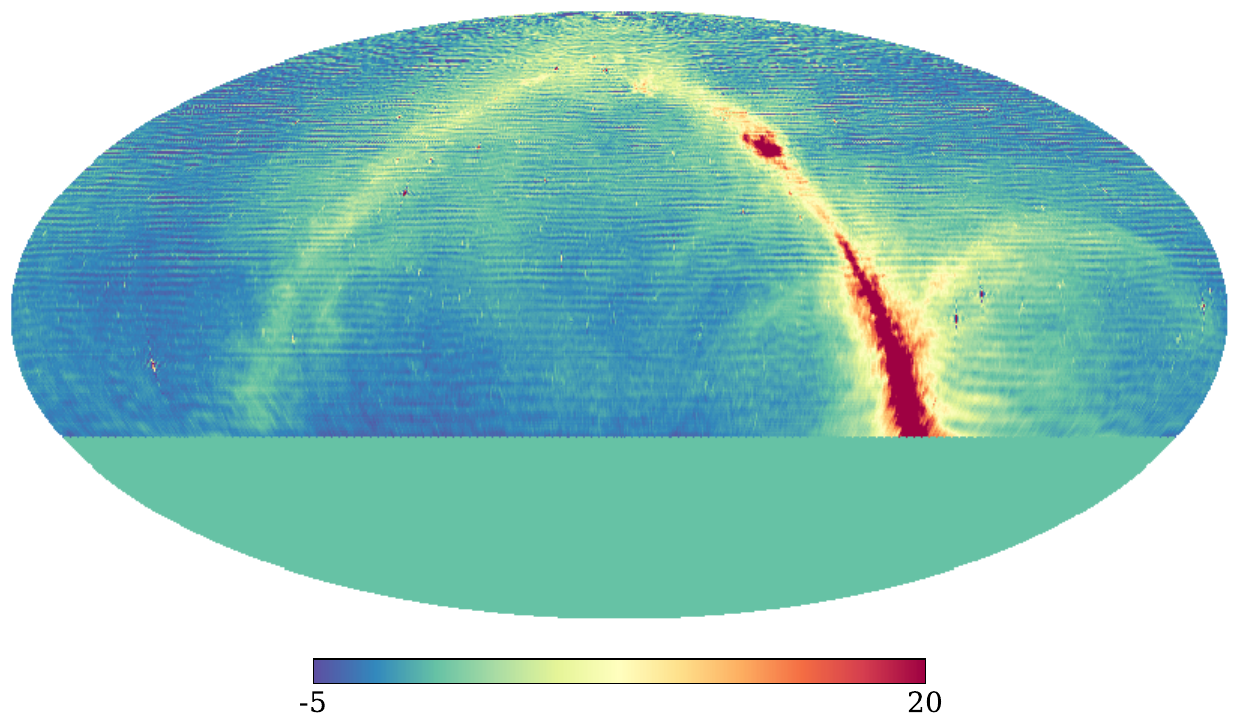}
    \includegraphics[width=0.33\textwidth]{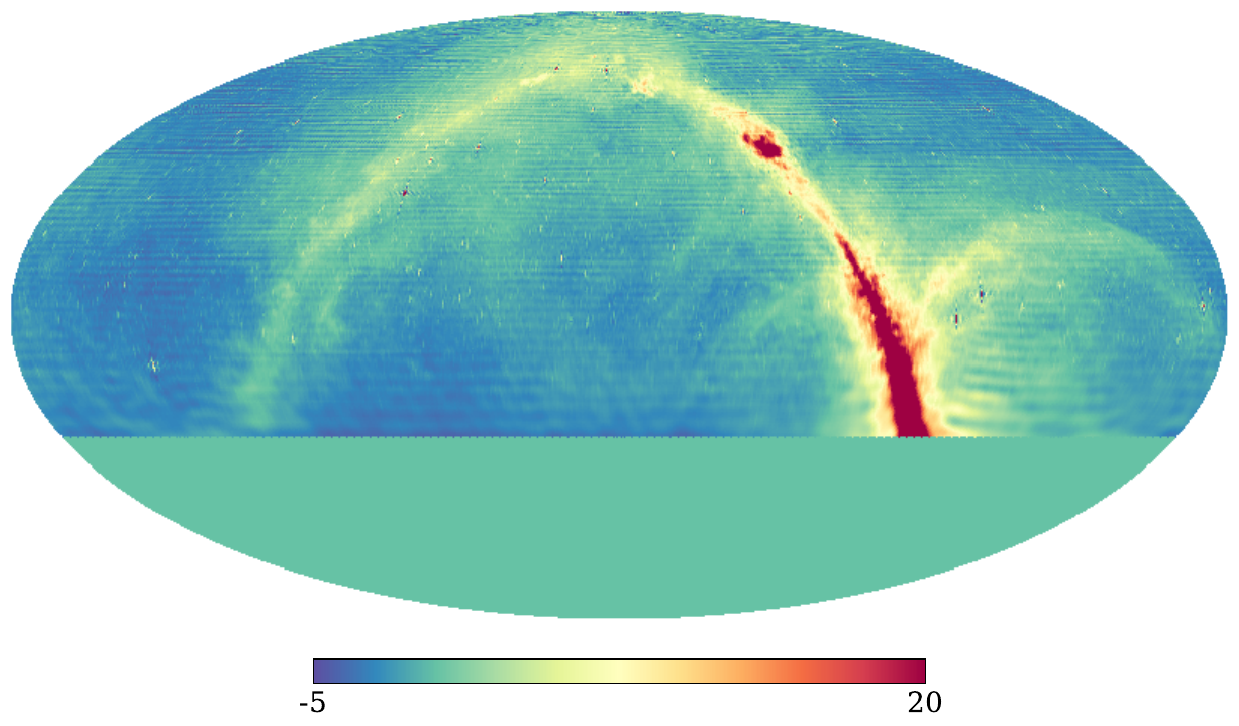}
    \includegraphics[width=0.33\textwidth]{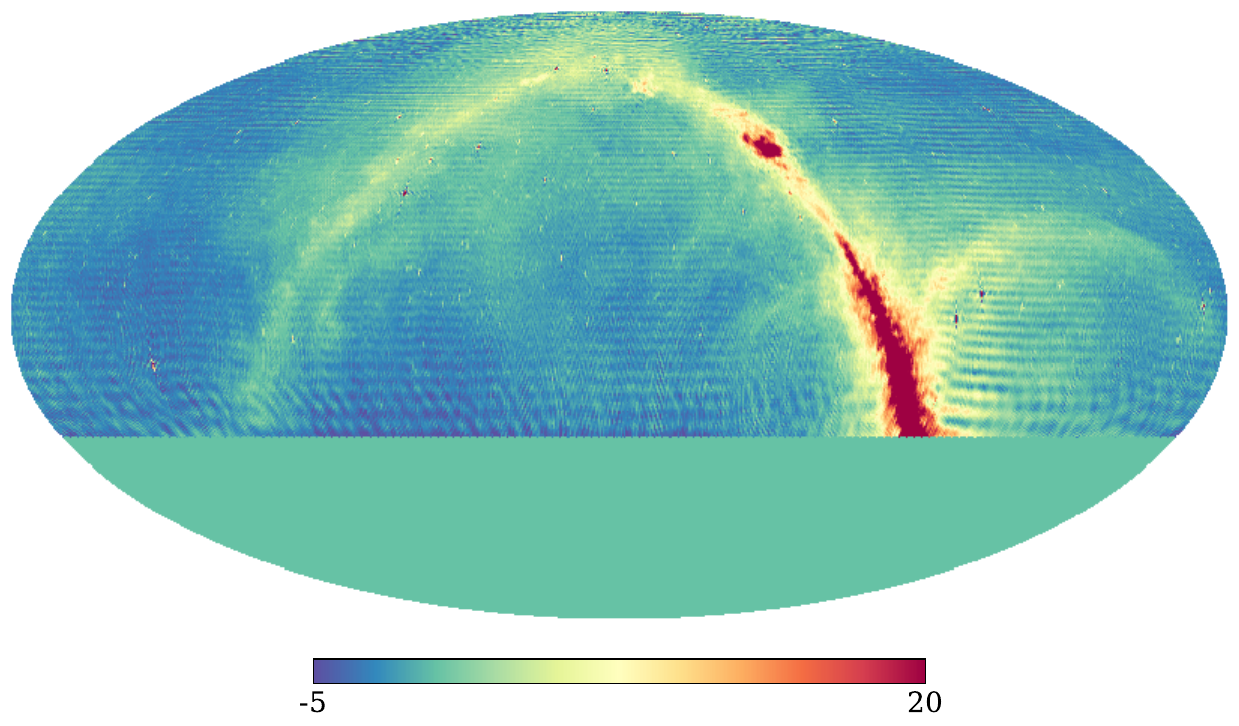}
    \includegraphics[width=0.33\textwidth]{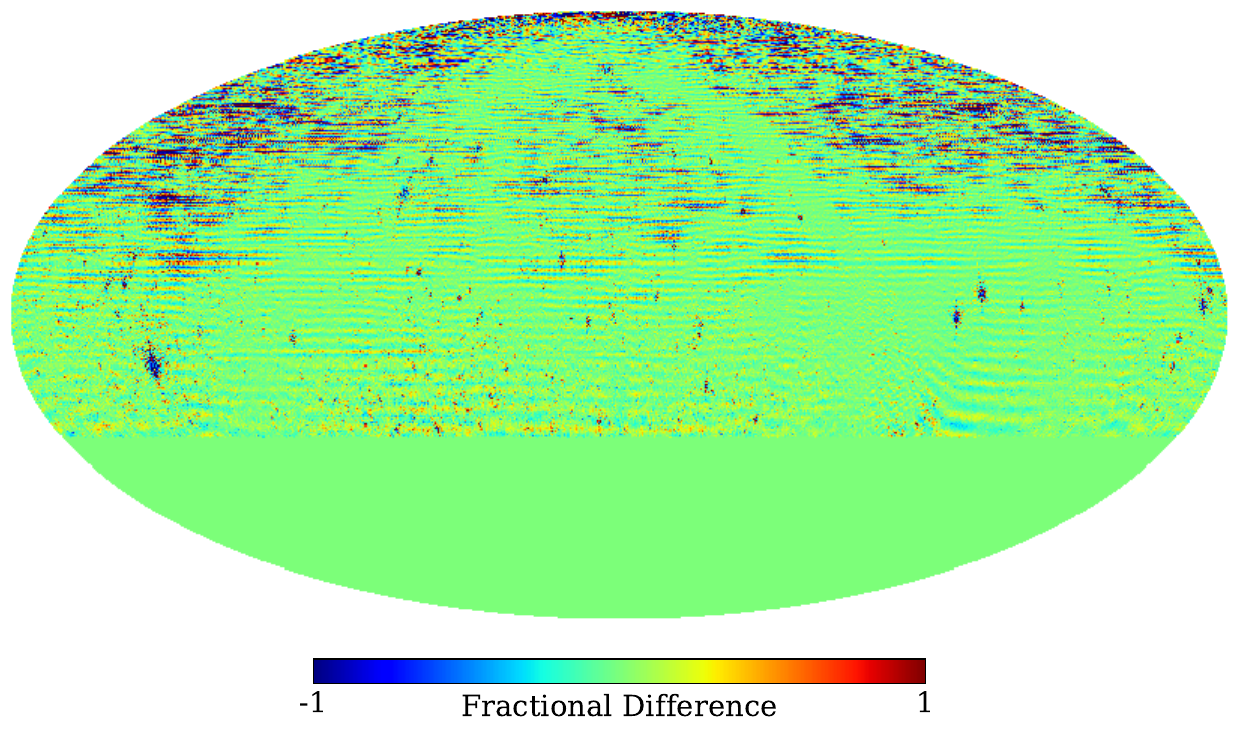}
    \includegraphics[width=0.33\textwidth]{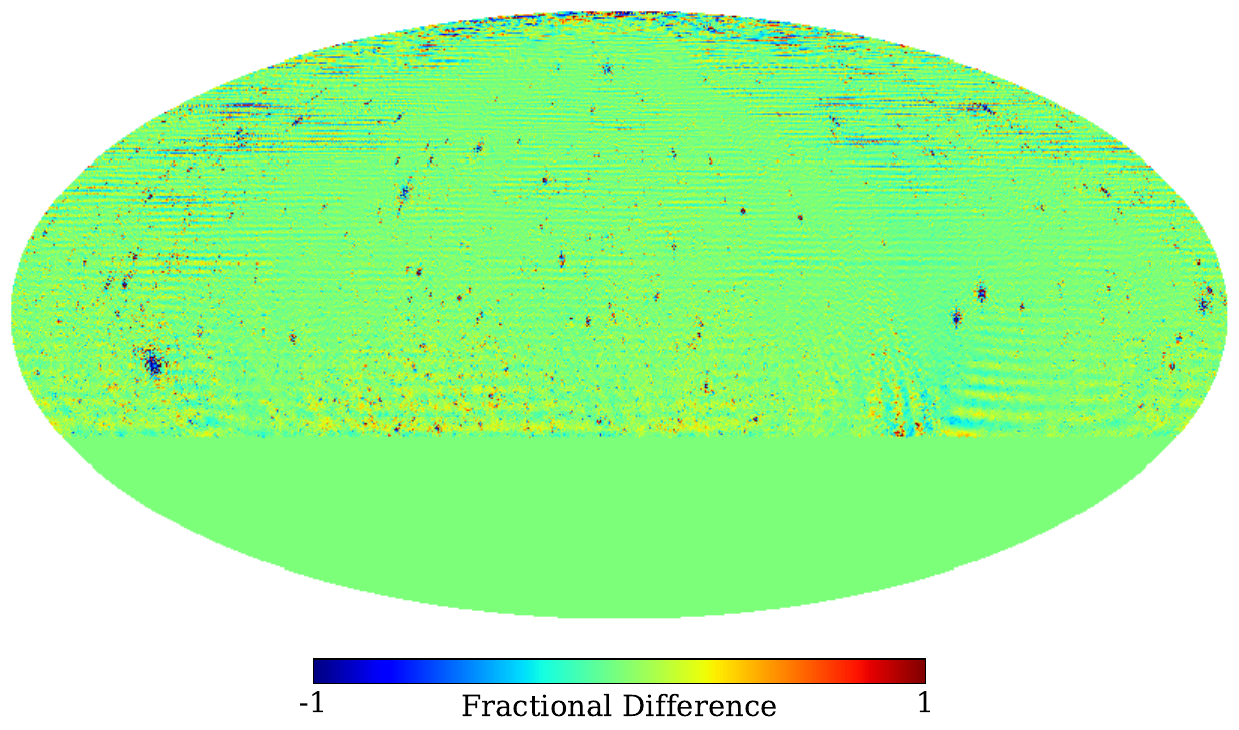}
    \includegraphics[width=0.33\textwidth]{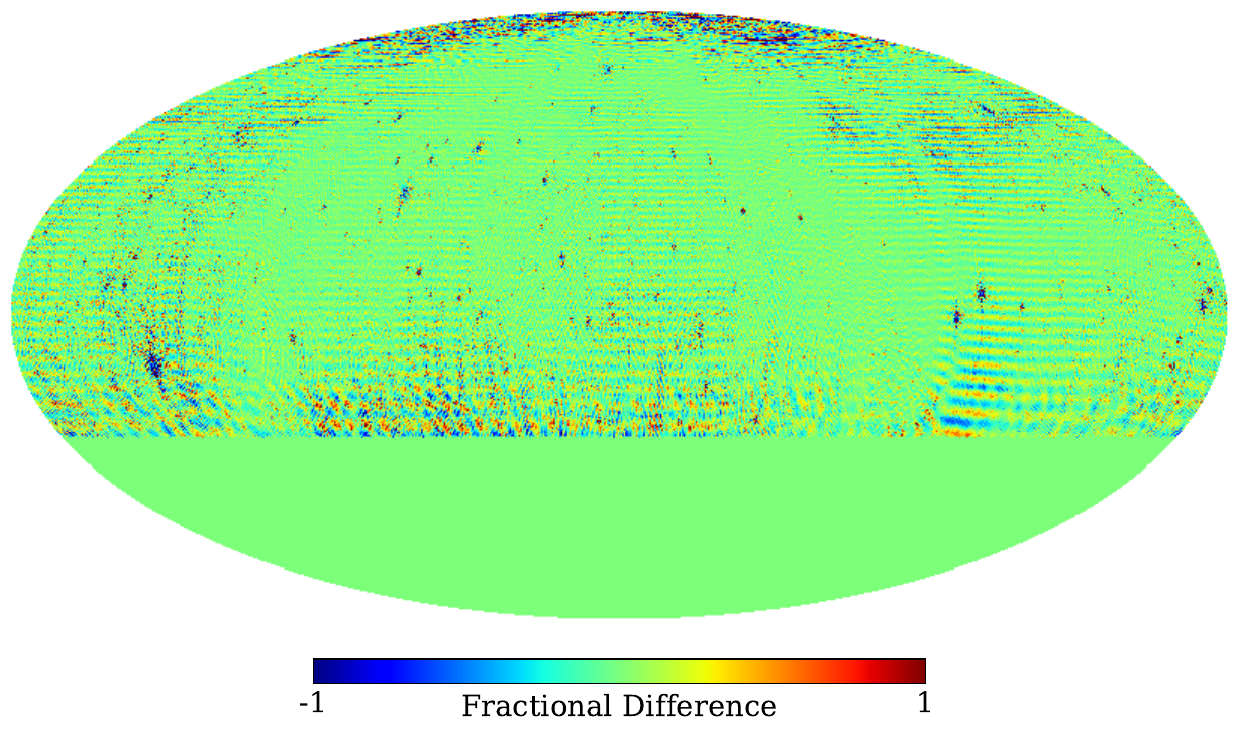}
    \caption{The reconstructed regularized maps (top) and their fractional error (bottom) using automatically determined regularization parameters. Left: Tikhonov regularization with $\lambda$ from the GCV method; Middle: Tikhonov regularization with $\lambda$ from the L-curve method; Right: TSVD regularization with threshold given by the L-curve method, all for the data with 30-day integration noise level. } 
    \label{fig:reconstruction-regularization-30day}
\end{figure}

In the top panels of Fig. \ref{fig:reconstruction-regularization-30day} we illustrate the reconstructed maps with the automatically chosen regularization parameters for the data with 30-day noise level, and in the bottom panels their relative error, i.e. the fractional difference with the input map. We show the cases of Tikhonov regularization using the GCV (left) and the L-curve (middle) criteria, and the TSVD regularization using the L-curve criterion (right). 

It seems that the Tikhonov regularized solutions generally yield visually better maps of the sky without the comb-like feature in the TSVD regularized map. The errors are typically small except around places where bright sources are located. With the $\lambda$ value obtained by the GCV criterion, which is smaller than the one chosen with the L-curve criterion, the region near the Galactic Center is better reproduced, but the northern region tends to be noisier. With the larger $\lambda$ value chosen using the L-curve criterion, the map is more smooth. However, we note that the $\lambda$ value chosen by both methods would vary with the noise level, therefore in the present case the larger $\lambda$ value obtained from the L-curve criterion produces a better overall visual impression, this may change for a different noise level or set up. Owing to the flat tail of the GCV function curve (Fig. \ref{fig:gcv}), the $\lambda$ value determined from the minimum of the GCV function might not always be accurate and satisfactory, while the L-curve method gives a more robust result. In the reconstructed maps, the smaller regularization parameter obtained from the GCV criterion leads to a more noisy reconstructed map compared to the one obtained using the L-curve criterion. As for the TSVD regularized map, it is generally similar to the map shown in Fig.~\ref{fig:re-map-tsvd-diff-reg-par}, and still shows the comb-like structure near the Galactic Center which is produced by the truncation of certain modes.

\subsection{An Overall Regularization Parameter}

As matrix $\vb*{B}$ in Equation \eqref{eq:mmode-3} is block diagonal, the $m$-mode method takes advantage of this structure by processing each $m$-block individually. To produce the regularized maps, we have also optimized the choice of regularization parameter for each $m$ separately, using the GCV and L-curve criteria. For simplicity, it is also possible to adopt a single overall regularization parameter $\lambda$ considering all $m$-blocks. We illustrate the L-curve criterion for Tikhonov regularization here. Then the point on the L-curve associated with the regularization parameter $\lambda$ is give by
\begin{equation*}
    \left(\sum_{m=0}^{m_{\rm max}}|| \vb*{B}_{m} \vb*{a}_{m}(\lambda) - \vb*{v}_{m}||_2, \quad \sum_{m=0}^{m_{\rm max}}|| \vb*{a}_{m}({\lambda})||_2 \right).
\end{equation*}

\begin{figure}
    \centering
     \includegraphics[width=0.45\textwidth]{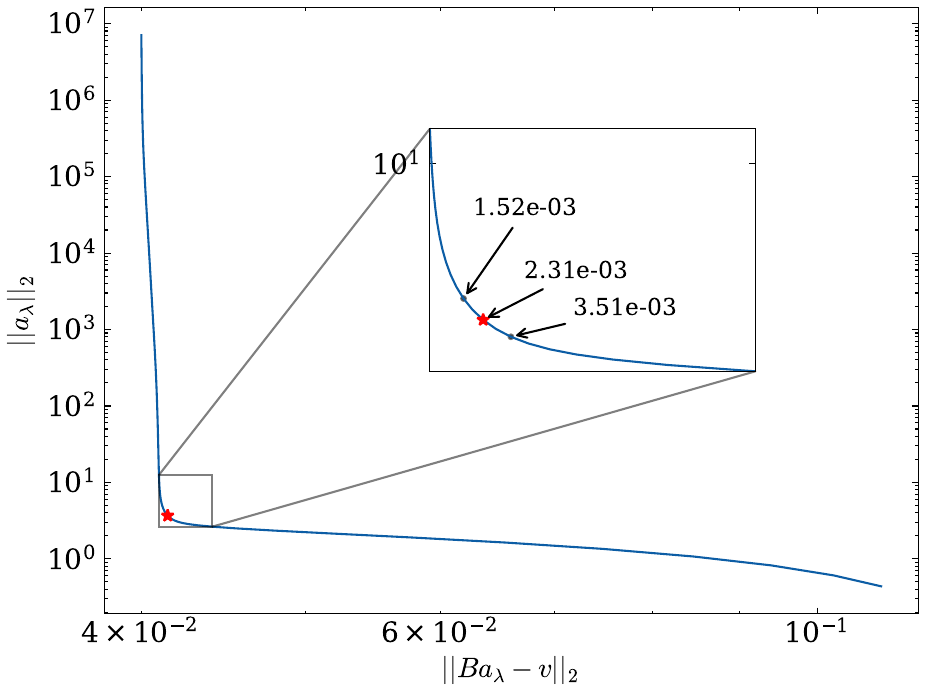}
    \includegraphics[width=0.45\textwidth]{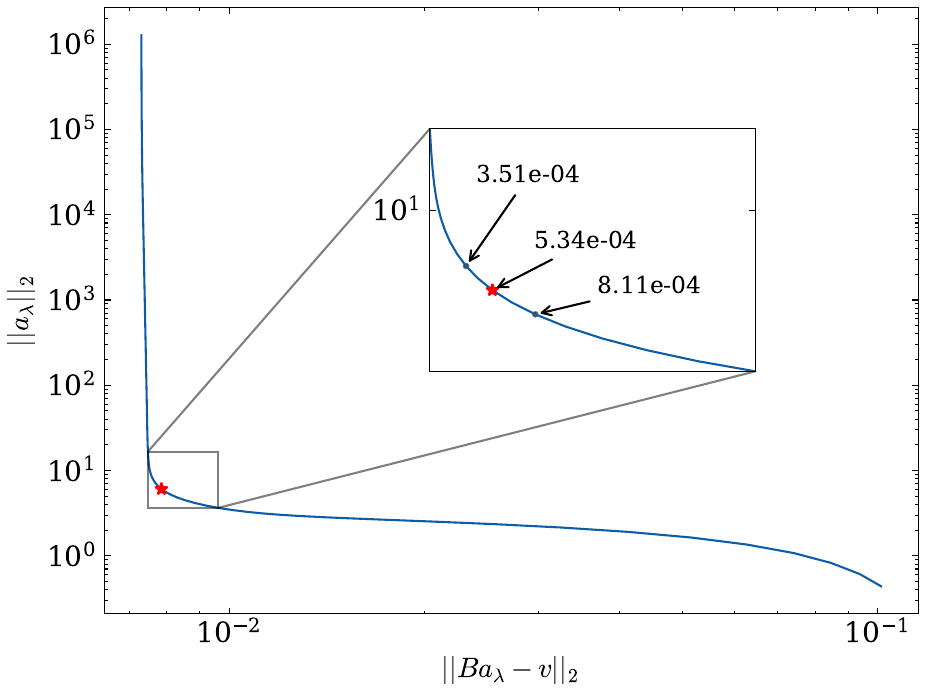}
    \caption{The overall (summing all $m$-modes) L-curve of Tikhonov regularization for the data containing 1-day (left) and 30-day integration noise (right), and red star marks the regularization parameter determined from each curve.}
    \label{fig:lcurve-one-without-prior}
\end{figure}

In Fig. \ref{fig:lcurve-one-without-prior}, we show the relevant L-curve for two cases with different noise level. For the 1-day case, the noise level is relatively high, and the L-curve criterion yields a relatively large optimal regularization parameter value $\lambda=2.31\times 10^{-3}$, while for the 30-day case, the level of noise is significantly lower, leading to a correspondingly smaller optimal regularization parameter value $\lambda=5.34\times 10^{-4}$.

\begin{figure}
    \centering  
    \includegraphics[width=0.45\textwidth]{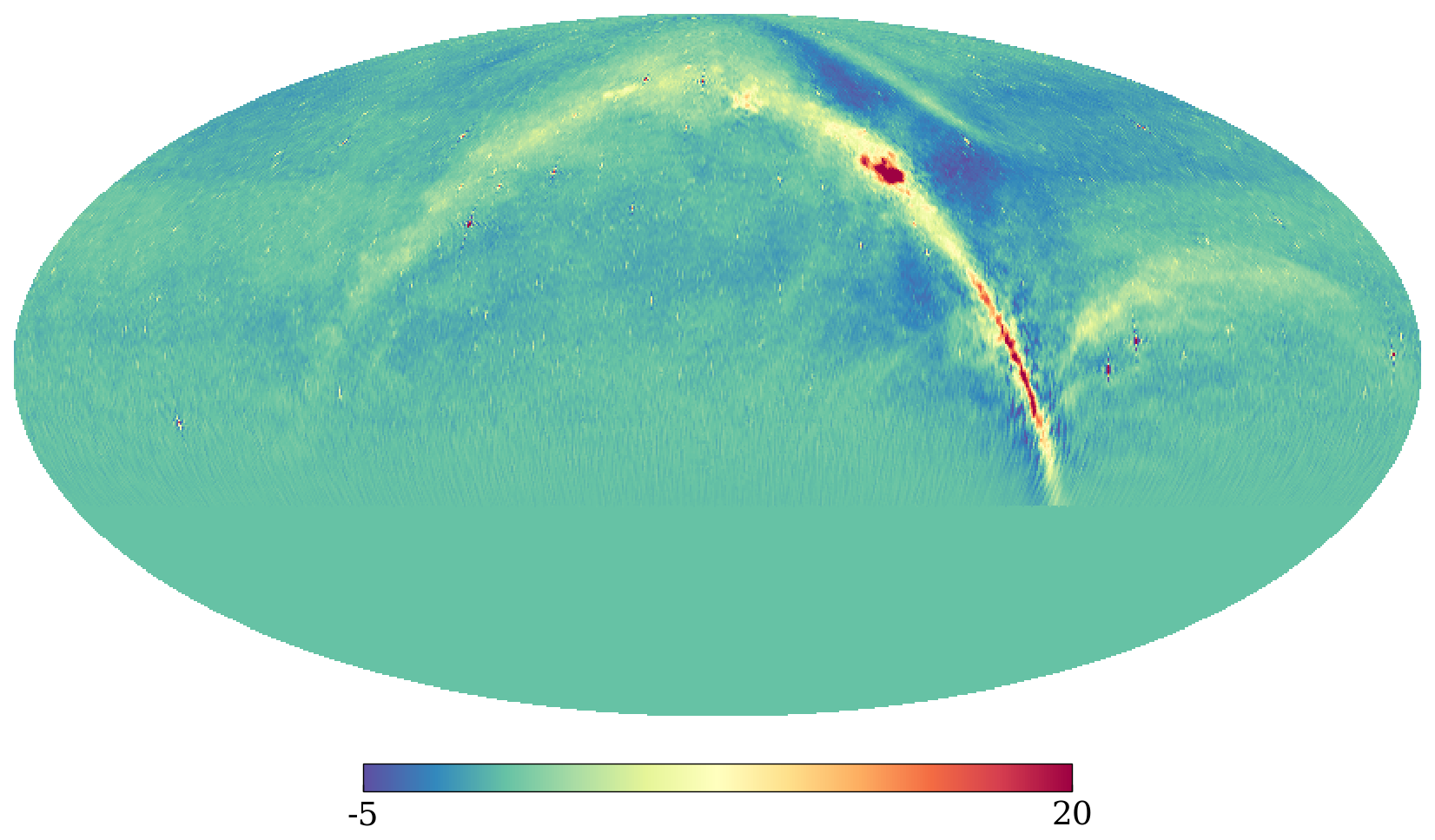}
    \includegraphics[width=0.45\textwidth]{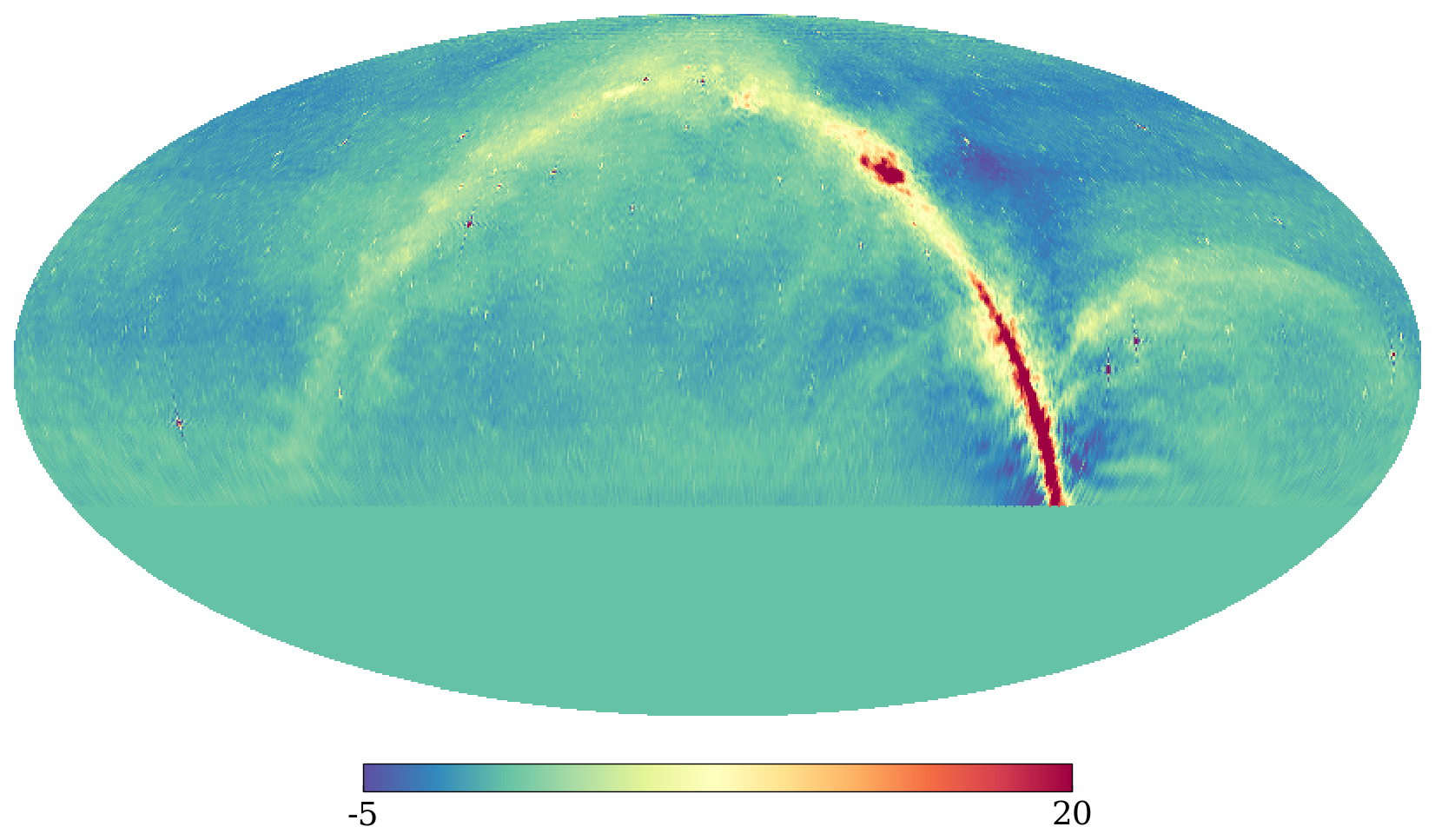}
    \caption{The Tikhonov regularized maps made with the optimal $\lambda$ determined from the overall L-curve for the data with 1-day (left) and 30-days noise (right). A larger regularization parameter (left) may result in an alias of the Galactic Center obviously in the map.}
    \label{fig:map_overall}
\end{figure}

In Fig.\ref{fig:map_overall} we illustrate the reconstructed map with these regularization parameters (the visibility data for imaging has the corresponding level of noise). In the 1-day case, due to the relatively large noise and regularization parameter, the map shows more deviation than the 30-day map, and the side lobe feature near the north celestial pole is quite obvious. However, we can see that for the 30-day noise level map, the deviation is also quite significant if we compare it to the maps made by the $m$-by-$m$ mode analysis shown in Fig.~\ref{fig:reconstruction-regularization-30day}. This is not surprising, since a single $\lambda$ value may not be suitable for all different $m$-modes. For those less sensitive modes that are susceptible, this value of $\lambda$ is perhaps too small to regularize the problem. On the other hand, for the modes to which the interferometer is sensitive, it may suppress too much to reconstruct the true signal.

\section{Discussions}
\label{sec:discussion}

\begin{figure}
    \centering
    \includegraphics[width=0.6\textwidth]{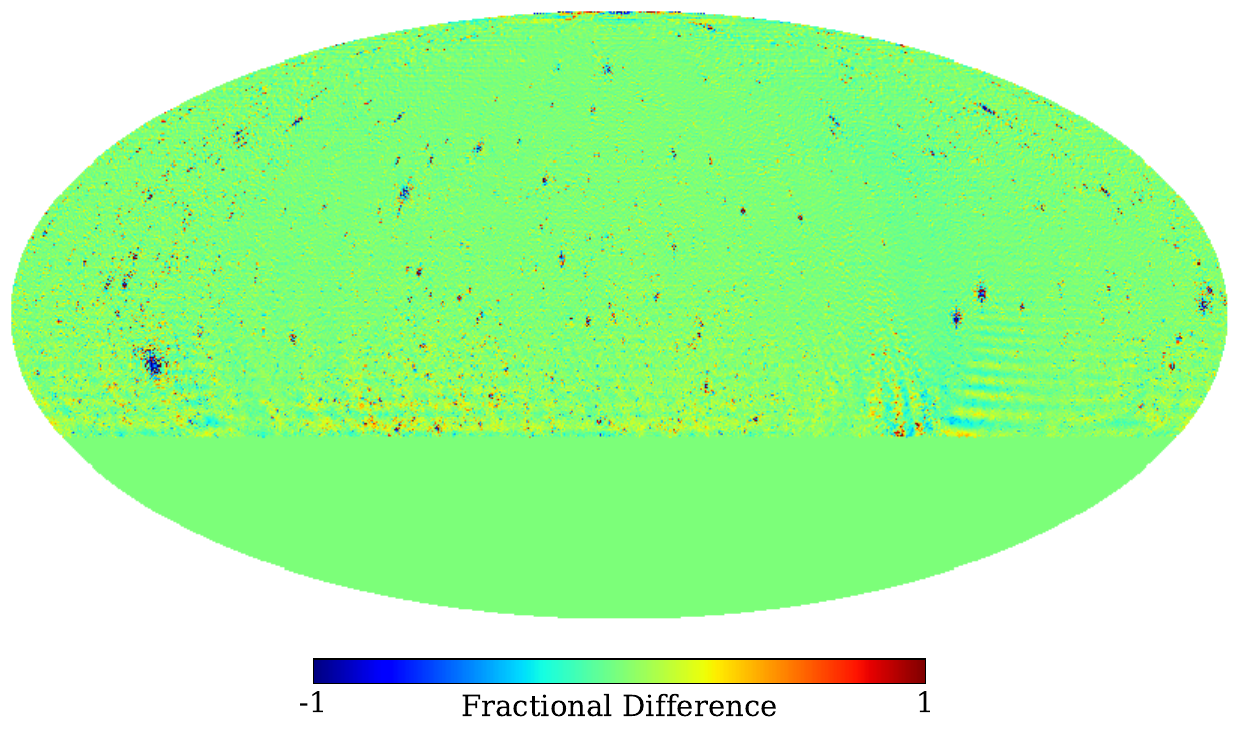}
    \caption{The fractional error caused by regularization in the Tikhonov regularized maps for the 30-day noise level data, where the regularization parameter is determined by the L-curve for each $m$. The overall effect of additional noise error is illustrated in Fig. \ref{fig:reconstruction-regularization-30day}. 
    }
    \label{fig:lcurve_30_bias_noise}
\end{figure}

Since the regularized solution is only an approximation of the true value, it inevitably introduces bias.
In Fig. \ref{fig:lcurve_30_bias_noise}, we show the fractional difference between the Tikhonov regularized map from the noise-free data and the input map, which gives the bias from regularization. We can see that the major bias is around the bright point sources and locates at the region near the horizon. The bias around the bright point sources is from their convolution with the point spread function, which is expect to be mitigated through further deconvolution techniques such as the CLEAN algorithm.

We can also quantify the quality of the reconstructed map by using the angular cross power spectrum between the input map and the reconstructed maps. In the cross power spectrum, defined as
\begin{equation*}
    C_{l}^{\rm re-in} = \left< a_{lm}^{\rm re} {a_{lm}^{\rm in}}^{*}\right> = \frac{1}{2l + 1}\sum^{l}_{m = -l} a_{lm}^{\rm re} {a_{lm}^{\rm in}}^{*},
\end{equation*}
the noise from the maps being crossed are uncorrelated, so they do not contribute to the cross power. 

\begin{figure}
    \centering
    \includegraphics[width=0.6\textwidth]{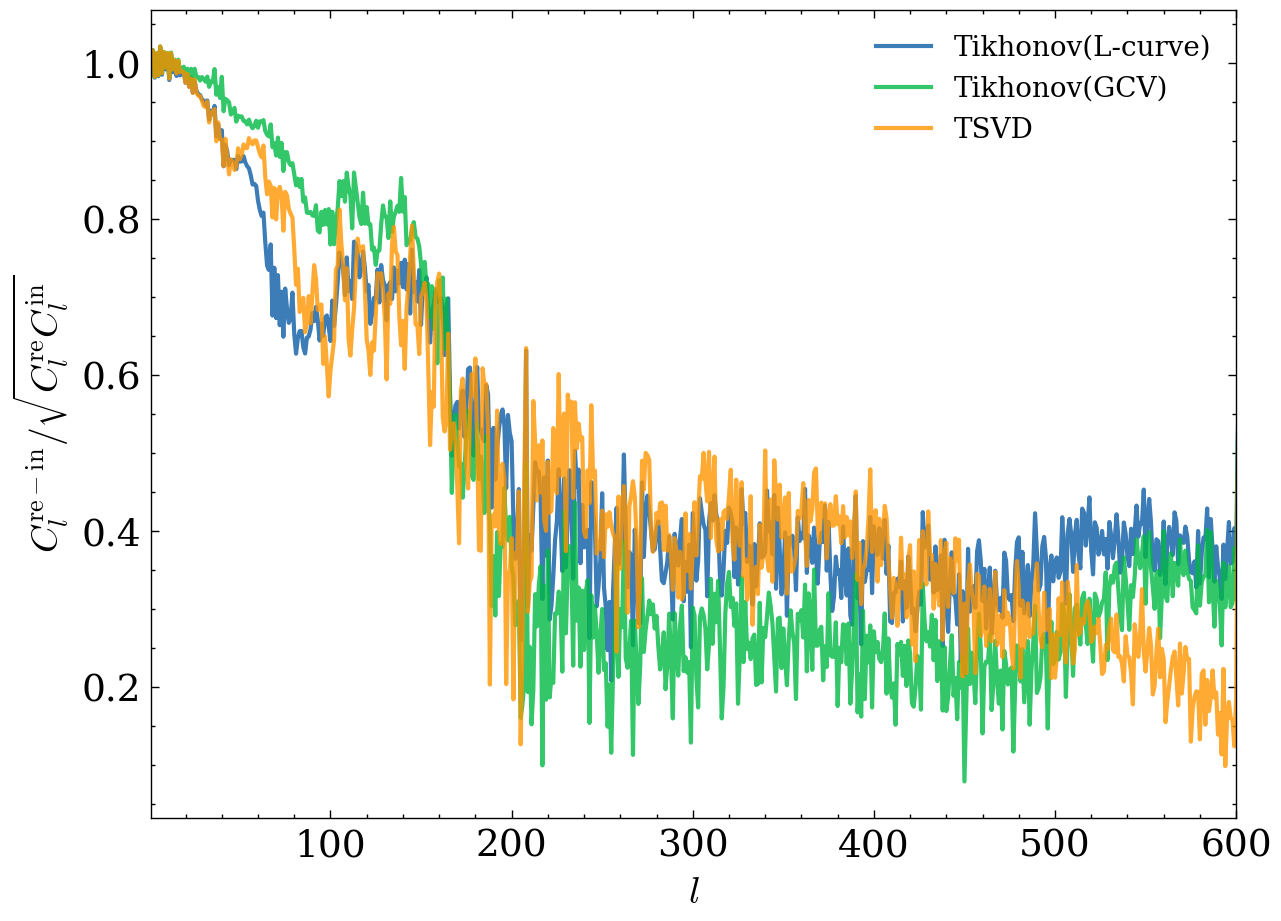}

    \caption{The cross-correlation coefficients$ C_l^{\rm re-in}/\sqrt{C_l^{\rm re} C_l^{\rm in}}$ for the Tikhonov regularized maps with the parameter determined using the GCV and L-curve criteria, and the TSVD regularized maps with the parameter determined using the L-curve criterion. All angular power spectra are computed for the observable part of sky ( $\delta \gtrsim -45^{\circ}$) and with the four bright point sources subtracted. 
    }
    \label{fig:angular_ps}
\end{figure}

In Fig. \ref{fig:angular_ps}, we show the cross-correlation coefficients, i.e. 
$ C_l^{\rm re-in}/\sqrt{C_l^{\rm re} C_l^{\rm in}}$, the ratio of the cross-correlation power between the reconstructed and input maps to the square root of the product of the corresponding auto  power spectra. For the Tikhonov regularization, we show the cases where the regularization parameter $\lambda$ is determined $m$-by-$m$ using the GCV and L-curve criteria, and for the TSVD regularization, the truncation threshold is determined based on the L-curve criterion. 

In all cases, the correlation falls below 1 as $l$ increases, which means that the correlation is not perfect due to reconstruction error. There is a general trend which all three curves follow. 
The result obtained from TSVD regularization has a correlation coefficient close to that of the Tikhonov regularization with the $\lambda$ determined by the L-curve at $l \leq 470$, but beyond this scale there is a significant deterioration. The result of the Tikhonov regularization with $\lambda$ determined by the GCV criterion exhibits higher correlation in the range $l \leq 150$, but shows lower correlation in the range $200 \leq l \leq 550$ than that  of the L-curve criterion,  where the noise is prone to be amplified. However, despite comparable performance as measured by the cross-correlation, the Tikhonov regularization produced better visual impression.

\section{Conclusion}
\label{sec:conclusion}

In this paper we investigated the regularization methods applied to the sky map reconstruction from the radio interferometer data taken by the Tianlai cylinder pathfinder array. Due to the incomplete $uv$-coverage from the limited number of baselines, regularization is generally necessary in the reconstruction of sky from interferometer data, but the strategy and method adopted differs case by case. In our previous paper \citep{2023RAA....23j5008Y}, we have investigated the map reconstruction for the Tianlai cylinder array, with emphasis on assessing the impact of array calibration error and noise on the final map, there we have only used the Moore-Penrose pseudo-inverse method in the map reconstruction. However, there are various different regularization methods, which also affect the map-making result. The exploration of the different regularization methods is the subject of the present paper. 

In this work we have studied the TSVD regularization and the Tikhonov regularization applied to our map reconstruction. The TSVD regularization is in some sense similar or equivalent to the Moore-Penrose pseudo-inverse used in paper I, which removes the modes susceptible to noise. The Tikhonov regularization, on the other hand, suppresses modes gradually, those susceptible to noise are more suppressed but not completely removed, consequently yielding a more smoothly regularized map. This smooth approach by the Tikhonov regularization can avoid the generation of obvious artefacts by the sharp cut off used in the TSVD regularization, producing visually better maps, though it is not necessarily more accurate than the TSVD regularization in a quantitative sense.

To obtain a high-fidelity sky map with the regularization techniques, it is crucial not to over-regularize the data by using an excessively large regularization parameter. However, the map may be greatly affected by the noise if a too small regularization parameter is chosen.
We have applied the GCV and L-curve methods to determine the optimal regularization parameter. We find that both methods do produce generally good maps for a reasonable level of noise. In our case the L-curve criterion provides a more stable regularization parameter, which also results in a map with better visual impression. However, we note that this result is specific to the case we investigated. For different data set, noise level, etc., the result can be different. Furthermore, although these methods can be used to optimize parameter selection, they are stilled based on simple reasoning. To achieve results that better meet specific requirements,  additional tuning may be needed.

\begin{acknowledgements}
We acknowledge the support of the National SKA program of China, No. 2022SKA0110100, the National Natural Science Foundation of China (Nos. 12203061, 12273070, 12303004).
\end{acknowledgements}



\bibliography{bibtex}
\bibliographystyle{raa}

\end{document}